\documentclass[%
 reprint,
superscriptaddress,
 amsmath,amssymb,
 aps,
]{revtex4-1}

\usepackage{graphicx}
\usepackage{dcolumn}
\usepackage{bm}
\usepackage{braket}
\usepackage[title]{appendix}
\usepackage[colorlinks,linkcolor=blue,citecolor=red,urlcolor=blue]{hyperref}
\usepackage{comment}

\usepackage{amsmath}

\usepackage{mathtools}

\usepackage{lipsum}

\begin{document}

\preprint{APS/123-QED}

\title{Electronic-nuclear entanglement in Born-Oppenheimer wave functions and beyond}
\author{Juan F. P. Mosquera}
\email{juan.pulgarin-mosquera@psi.ch}
\thanks{Corresponding author}
\affiliation{PSI Center for Scientific Computing, Theory and Data, 5232 Villigen PSI, Switzerland}
\affiliation{Department of Physics, University of Fribourg, CH-1700 Fribourg, Switzerland}
\author{Jos\'e Luis Sanz-Vicario}%
\email{jose.sanz@udea.edu.co}
\thanks{Corresponding author}
\affiliation{Grupo de Física Atómica y Molecular, Instituto de Física, Universidad de Antioquia, Medellín, Colombia}

\date{\today}

\begin{abstract}

We analyze the entanglement between electronic and nuclear motions in molecular wave functions widely used by theoretical chemists; namely, i) Born-Oppenheimer separation in the adiabatic picture, ii) the transformation into a diabatic picture, iii) using a Born-Huang expansion, and iv) the eigenfunction of the full molecular Hamiltonian. Our showcase is based on two one-electron one-dimensional molecular Hamiltonians (H$_2^+$ and the Shin-Metiu model). 
We find that within the Born-Oppenheimer approximation, any molecular state (although separable) is always entangled, and its entanglement content may be assessed by the variation of the electronic wave function along the different nuclear geometries, with the nuclear wave function indeed playing the role of a tester. The presence of avoided crossings among the adiabatic potential energy curves brings about dramatic changes in the entanglement content of the wave function: sharp avoided crossings favor a diabatic picture (real crossings between potential energy curves) while in broad avoided crossings the adiabatic picture prevails. The total eigenfunction of the molecular Shin-Metiu Hamiltonian indicates that nuclear densities accommodate well within the diabatic curves for strong adiabatic couplings but within adiabatic curves for weak ones. Consequently, we find that the electron-nuclei entanglement content is a valid witness to unveil strong or weak non-adiabatic couplings in molecules.
In terms of entanglement, we also find that the Born-Huang expansion, based on Born-Oppenheimer adiabatic electronic states, does not provide a correct trend of entanglement compared with that of the total molecular eigenfunction, thus indicating a very slow convergence of this
expansion.

\end{abstract}

\maketitle

\section{\label{sec:intro}Introduction}

The Born-Oppenheimer (BO) approximation in terms of the small (large) ratio between electron and nuclear masses (velocities) has been the preamble \cite{Born1927} for the widely used adiabatic separation between electron and nuclear motion in the quantum molecular problem \cite{Born1954, Longuet-Higgins1961, Bransden2003}. 
The corresponding BO wave function (WF) is an approximation to the exact molecular WF, namely $\Phi_{\mathrm{Mol}} ({\bf x},{\bf R}) \sim \phi^{\mathrm{BO}}({\bf x};{\bf R}) \chi^{\mathrm{BO}}({\bf R})$, where the electronic WF $\phi^{BO}( {\bf x};{\bf R} )$ is obtained as an eigenstate of the electronic Hamiltonian for a chosen clamped nuclear geometry ${\bf R}$. 
This separation has been championed by quantum chemists for decades because it allows the concept of a nuclear motion (with a WF $\chi^{BO} ({\bf R})$) driven by unambiguous potential energy surfaces provided by the electronic solution made a priori.

Other separations are possible, even an exact factorization in a similar form; $\Phi_{\mathrm{Mol}} ({\bf x}, {\bf R}) = \phi({\bf x}|{\bf R}) \chi({\bf R})$, but as a conditional probability amplitude for the electronic part, as stated by the Bayes theorem \cite{Hunter1975}. This proposal has been further investigated by other authors \cite{Bishop1975,Czub1978,Gidopoulos2014, Cederbaum2013} to find that the reduction to conditional probability amplitude requires knowledge of the total WF, that the equations governing the new two factorized WFs are much more complicated to solve, and that the non-adiabatic potentials depend upon each rovibrational state. 
To our knowledge, the time-independent exact factorization has been applied to a limited number of soluble model systems \cite{Cederbaum2013}, and the prospect of its usability in other complex molecular systems remains doubtful. Whatever the case, despite an extant exact factorization with a difficult manipulation, the molecular state $\phi({\bf x}|{\bf R}) \chi({\bf R})$ still remains entangled. However, this kind of factorized solution, unfamiliar to most quantum chemists, is not within the scope of the present work.

The BO separation also provides a particular protocol for approaching the full solution of the total Hamiltonian, by proposing the BO electronic states as a complete basis (see \cite{Sutcliffe2010,Sutcliffe2012} for a historical account and a critical viewpoint on solutions based on the BO adiabatic approximation) in the so-called Born-Huang (BH) expansion \cite{Born1954}.
This BH expansion is not an strictly exact solution and it introduces non-adiabatic couplings between the BO adiabatic electronic states, which partially dilutes the concept of nuclei moving on a single BO adiabatic potential energy surface. 
The accurate computation of non-adiabatic couplings near avoided crossings or conical intersections (at geometries where the BO fails) is usually a remarkably complicated task that is avoided by the proposal of unitary transformations to a diabatic picture \cite{Smith1969}. 
The diabatization procedure eliminates non-adiabatic couplings in favor of softer electrostatic couplings, since the diabatic WFs are no longer eigenstates of the electronic Hamiltonian. 

In this work, we deal with approximate BO WFs in the adiabatic picture and their transformation into diabatic WFs, as well as beyond the BO approximation, with a BH expansion in terms of adiabatic BO WFs and finally by using a direct variational ansatz to solve the total molecular Hamiltonian. 
The entanglement between electron and nuclear motions has already been analyzed by several authors \cite{Bouvrie2014, McKemmish2015,Izmaylov2017,Sanz2017} and also from a time-dependent perspective \cite{Vatasescu2013}, including the use of the time-dependent exact factorization ansatz \cite{Agostini2019}.
In our previous work \cite{Sanz2017}, we computed the electro-nuclear entanglement of the simplest molecule H$^+_2$ from first principles, without considering BO or Born-Huang ansatzes. 
Without bias, we diagonalize the total molecular Hamiltonian represented with a basis of tensor products of crude electron and nuclear basis functions $\Phi({\bf x},{\bf R}) = \sum C_{nm} \phi_n({\bf x}) \chi_m({\bf R})$. 
With this method of solution beyond BO, we obtained a variational solution for all vibronic states up to a given energy, devoid of any reference to BO potential energy curves. This kind of solution requires large expansions and is computationally very demanding for large molecules. However, this approach can be fairly implemented for 1D molecular models. The total variational WF is devoid of any semiclassical BO clamped nuclei approximation and much closer to the exact total WF, thus serving us as a reference.  


The advantage of using molecular models of reduced dimensionality (for H$^+_2$ and the Shin-Metiu molecular model) is that both electronic and nuclear WFs are accessible and manageable in computational terms, and can be easily visualized. 
Also, non-adiabatic couplings at avoided crossings can be readily computed with almost exact variational electronic WFs. 
Therefore, although we analyze electro-nuclear entanglement in these simple models, they include all the ingredients already present in the molecular dynamics involving excited states of complex molecules. 

The correlation, coherence, and entanglement between particles within a whole molecular system are ongoing continuous objects of study from different perspectives. 
Correlation among particles are fully imprinted in the Hamiltonian, and its proper description depends upon the chosen ansatz for the WFs, the latter carrying the characters of coherence and entanglement. 
For example, the particular electron-nuclei entanglement in the photoionization of H$_2^+$ subject to two delayed attosecond pulses has recently been theoretically and experimentally investigated \cite{Vrakking2021,vrakking2022}. 
A desirable future goal is to understand how to take control on the content of entanglement between particles or motions within complex systems, and to take advantage of it in building efficient entanglement-based quantum-key distribution systems.

The paper is organized as follows. We introduce the theory in Section \ref{sec:theory}: BO states, BH expansion, non-adiabatic couplings, the application of the Schmidt theorem, and the calculation of von Neumann entropies. 
Both electronic and nuclear problems are solved numerically in a 1D grid using a Fourier grid Hamiltonian method.
In Section \ref{sec:results} we first analyze the molecular case of 1D H$^+_2$ (without avoided crossings between the lowest excited states), and introduce a simplified procedure to assess the entanglement using a much reduced representation of the nuclear density matrix.
Secondly, the Shin-Metiu model serves as an excellent platform to analyze entanglement with BO adiabatic vs. diabatic pictures, and with BH expansion, all compared to the entanglement content in the variational WF of the total molecular Hamiltonian. 
We end up with conclusions and perspectives. Atomic units are used unless otherwise stated. 

\section{ Theory}
\label{sec:theory}
\subsection{\label{sec:BOA} Born-Oppenheimer approximation in molecules}

In molecular systems, composed of both electrons and nuclei, a natural bipartite arises between electronic and nuclear degrees of freedom. The total molecular Hamiltonian can be written as
\begin{equation}
    \hat{H} = \hat{T}_e + \hat{T}_N + \hat{U}(\mathbf{x}, \mathbf{R}),
    \label{eq:molecular_hamiltonian}
\end{equation}
where $\hat{T}_e$ and $\hat{T}_N$ are the kinetic energy operators for electrons and nuclei, respectively, and $\hat{U}$ encompasses all Coulomb interactions: electron-electron, nuclei-nuclei, and electron-nuclei. 
The collective electronic coordinates are denoted by $\mathbf{x}$ and the nuclear coordinates by $\mathbf{R}$. 
The BO approximation is based on the observation that nuclei are much heavier and thus move more slowly than electrons. As a result, one can treat the nuclei with a fixed geometry while solving the electronic problem. This allows the total molecular WF to be approximated as a product of electronic and nuclear components, that reads
\begin{equation}
    \Psi^{\text{BO}}_{n} (\mathbf{x}, \mathbf{R}) = \phi_n (\mathbf{x}; \mathbf{R}) \chi_{n,m}(\mathbf{R}),
    \label{eq:Born-Oppenheimer_wave_function}
\end{equation}
where $\phi_n(\mathbf{x}; \mathbf{R})$ is the electronic WF parameterized by the nuclear coordinates and $\chi_{n,m}(\mathbf{R})$ the nuclear WF for the $n$-th electronic state and $m$-th nuclear state.
To determine the electronic WF and energy, one solves the electronic Schrödinger equation in each fixed nuclear geometry $\mathbf{R}$, i.e.
\begin{equation}
    \left[ \hat{T}_e + \hat{U}(\mathbf{x}, \mathbf{R}) \right] \phi_n(\mathbf{x}; \mathbf{R}) = E_n(\mathbf{R}) \phi_n(\mathbf{x}; \mathbf{R}),
    \label{eq:electronic_hamiltonian}
\end{equation}
where $E_n(\mathbf{R})$ defines the potential energy surface (PES) associated with the $n$-th electronic state. These surfaces serve as effective potentials for the nuclear motion, governed by the respective nuclear Schrödinger equation
\begin{equation}
    \left[ \hat{T}_N + E_n(\mathbf{R}) \right] \chi_{n,m}(\mathbf{R}) = W^n_m \chi_{n,m}(\mathbf{R}),
    \label{eq:eq:nuclear_hamiltonian}
\end{equation}
with $W^n_m$ being the total vibronic energy of the $m$-th nuclear state bound by the $n$-th PES.

The BO approximation is widely used in molecular physics and quantum chemistry because of its computational simplicity and its ability to yield accurate results when non-adiabatic effects can be safely neglected. Moreover, this approximation provides a valuable insight and serves as a foundation for more general solutions, as described in the next section.

\subsection{\label{sec:born-huang-hamiltonian} Beyond BO: the Born-Huang ansatz and the variational ansatz.}
To account for non-adiabatic effects and going beyond the standard BO approximation, one can use a more general ansatz for the total electro-nuclear WF, given by
\begin{equation}
    \Psi^{\text{BH}}(\mathbf{x},\mathbf{R}) = \sum_n f_n(\mathbf{R}) \phi_n(\mathbf{x};\mathbf{R}),
\end{equation}
named Born-Huang (BH) expansion, where $\phi_n(\mathbf{x};\mathbf{R})$ is the complete set of adiabatic BO electronic eigenfunctions in Eq.~\eqref{eq:electronic_hamiltonian}, and $f_n(\mathbf{R})$ are the nuclear coefficients that carry the $\mathbf{R}$-dependence in the total state \cite{Born1954,Sutcliffe2010,Barford2013}.
Inserting this BH ansatz into the total molecular Hamiltonian [Eq.~\eqref{eq:molecular_hamiltonian}] and projecting with $\int d\mathbf{x} \,\phi_{n'}^*$, it yields the following set of coupled equations for the nuclear coefficients
\begin{align}
\label{eq:BHnuclear}
    & \left( \hat{T}_N +  E_n(\mathbf{R}) -  W^{\text{BH}}_n \right)  f_n(\mathbf{R}) \nonumber \\
    & \qquad \qquad + \sum_{n'}\hat{C}_{n,n'}( \mathbf{R},\hat{\mathbf{P}})f_{n'}(\mathbf{R})=0,
\end{align}
with $W^{\mathrm{BH}}_n$ being the total BH energy, and $\hat{C}_{n,n'}(\mathbf{R}, \hat{\mathbf{P}})$ corresponds to the non-adiabatic coupling operator between electronic states $n$ and $n'$, This coupling operator is Hermitian and takes the well-known form
\begin{equation}
    \hat{C}_{n,n'}(\mathbf{R}, \hat{\mathbf{P}}) = \frac{1}{2\mu_M}\Big( 2 \mathbf{A}_{n,n'}(\mathbf{R}) \cdot \hat{\mathbf{P}} + B_{n,n'}(\mathbf{R}) \Big),
    \label{eq:non-adiabatic-couplings}
\end{equation}
where $\mathbf{A}_{n,n'}$ and $B_{n,n'}$ include the first- and second-order coupling terms, and $\mu_M$ is the reduced mass of the nuclei. These electronic non-adiabatic couplings are represented, for example, for diatomic molecules as
\begin{equation}
 {\bf A}_{n,n'} (R) = \int d{\bf x} 
\phi_n ({\bf x};{R}) \frac{\partial}{\partial R}\phi_{n'} ({\bf x}; R) 
\end{equation}
\begin{equation}
{ B}_{n,n'} ({\bf R}) = \int d{\bf x} 
\phi_n ({\bf x};{R}) \frac{\partial^2}{\partial R^2}\phi_{n'} ({\bf x};R).  
\end{equation}
The vector coupling matrix elements ${\bf A}_{n,n'}$ are usually computed pseudo-analytically or numerically within some molecular computational toolkits, but the second scalar coupling $B_{n,n'}$ is usually ignored.
This BH approach provides a convenient representation for evaluating observables, population transfer, and coherence effects in a fully coupled electro–nuclear picture. As such, it forms the foundation for many modern techniques in the treatment of photo-excitation, proton-coupled electron transfer, and other non-adiabatic collisional phenomena \cite{Nelson2020, Song2021, Bransden1992}.

\subsubsection{\label{sec:BH-BO-grid}\bf 
Matrix representation in a nuclear basis}

To solve the coupled Eqs.~\eqref{eq:BHnuclear} numerically, one can expand the nuclear coefficients in terms of the BO nuclear eigenfunctions $\{\chi_{n,m}(\mathbf{R})\}$,
\begin{equation}
    f_n(\mathbf{R}) = \sum_m \Lambda_{nm} \chi_{n,m}(\mathbf{R}),
\end{equation}
where $\Lambda_{nm}$ are the expansion coefficients. This leads to the following matrix eigenvalue problem,
\begin{equation}
    (W^n_m - W^{\mathrm{BH}}_n) \Lambda_{nm} + \sum_{n'm'} C_{n,n',m,m'} \Lambda_{n'm'} = 0,
\end{equation}
being $C_{n,n',m,m'}$ the vibronic non-adiabatic coupling matrix elements. Finally, the total electro-nuclear WF is then expanded as
\begin{equation}
    \Psi^{\text{BH}}(\mathbf{x},\mathbf{R}) = \sum_n \phi_n(\mathbf{x};\mathbf{R}) \sum_m \Lambda_{nm} \chi_{n,m}(\mathbf{R}).
    \label{eq:Born-Huang_wave_function}
\end{equation}
In case all non-adiabatic couplings are neglected, the expansion reduces to a single BO term.

\subsubsection{\label{sec:fullHam}\bf 
Variational ansatz for the total wave function.}

A direct variational solution of the total molecular Hamiltonian in Eq.~ \eqref{eq:molecular_hamiltonian} can also be implemented by using a configuration interaction tensor product, with separate orthogonal basis functions for electrons and nuclei to expand each molecular state, i.e.,
\begin{equation}
\label{eq:totalvariationalWF}
\Psi ({\bf x},{\bf R})=\sum_{mn} C_{m,n} \phi_n({\bf x}) \chi_m({\bf R}),
\end{equation}
where the expansion coefficients $C_{m,n}$ are obtained computationally after diagonalization. Of course, this approach can be implemented in practice for Hamiltonians with a reduced number of degrees of freedom.

\subsection{\label{sec:Schmidt_theorem}Schmidt theorem for bipartite systems}

The structure of molecular WFs under the BO, the BH or the full variational ansatz naturally leads to a bipartite quantum system for the electronic and nuclear halfspaces. For this case, the singular value decomposition (SVD) provides a complete representation in terms of two subsystems, $U$ and $V$. 
This SVD-based representation is mathematically equivalent to the Schmidt theorem \cite{Gerry2004, Sanz2017,Blavier2022}. Therefore, employing the Schmidt decomposition with all singular values provides an exact description of the quantum state, although in practice only a limited number of separable terms are often required to achieve an accurate approximation. 

\subsubsection{\textbf{Schmidt theorem for a pure quantum state}}
Given a pure state $\left| \Psi \right\rangle$ of a bipartite system ($U$,$V$) with Hilbert space $\mathcal{H} = \mathcal{H}_U \oplus \mathcal{H}_V$, the Schmidt theorem confirms the existence of new orthonormal sets $\left\{ u_1, \dots, u_M \right\} \in \mathcal{H}_U$ and $\left\{ v_1, \dots, v_N \right\} \in \mathcal{H}_V$ such that
\begin{equation}
    \left| \Psi\right> = \sum_{i=1}^{\mathrm{min}(M,N)}{\sqrt{\lambda_i} \left| u_i \right> \otimes \left| v_i \right>},
    \label{eq:schmidt_decomposition}
\end{equation}
with $\left| u_i \right>$ and $\left| v_i \right>$ being the reduced Schmidt bases, obtained as eigenstates of the reduced density matrices in the half-spaces, namely $\hat{\rho}^U = \text{Tr}_{V}\left[\hat{\rho} \right]$ and $\hat{\rho}^V = \text{Tr}_{U}\left[\hat{\rho} \right]$. 
It should be noted that both reduced operators share the same eigenvalue spectrum $\{\lambda_i\}$.
These eigenvalues satisfy $0 \leq \lambda_i \leq 1$ and $\sum_{i=1}^{\min(M,N)} \lambda_i = 1$, as expected for a normalized pure state. For the indices $\min(M,N) < i \leq \max(M,N)$, the Schmidt eigenvalues $\lambda_i$ and the expansion coefficients $\sqrt{\lambda_i}$ are identically zero. In this work, we are using a coordinate grid representation for both the electronic and nuclear half-spaces. Then $\{x_1,...,x_N\} \in \mathcal{H}_{el}$ and $\{R_1,...,R_M\} \in \mathcal{H}_{nuc}$ and the Schmidt basis of the eigenstates are thus represented as $u_i(x_j) = \langle x_j |u_i \rangle$ and $v_i(R_j) = \langle R_j | v_i \rangle$.  

\subsubsection{\textbf{Entanglement measure}}

The degree of entanglement can be quantified by the von Neumann entropy, which is defined as
\begin{equation}
    S=-\text{Tr}\left[\hat{\rho}^x \log{\hat{\rho}^x } \right]=-\sum_i{\lambda_i\log{\lambda_i}},
    \label{eq:entropy}
\end{equation}
where $x$ stands for $U$ or $V$.  When the expansion coefficient in Eq.~\eqref{eq:schmidt_decomposition} satisfies $\sqrt{\lambda_i} = \delta_{ij}$ the expansion reduces to a single product state term with a single eigenvalue $\lambda=1$ (that is, without entanglement) and the entropy $S$ is clearly zero. As the entanglement increases, so does the entropy, with larger values indicating substantial quantum correlations between subsystems in this particular half-space separation.

\subsubsection{\textbf{Wave functions in the Schmidt form}}

The BO, BH and the variational total wave functions may eventually be separable in electronic and nuclear coordinates, although they cannot be recast into a product state in the form $\Psi(\mathbf{x},\mathbf{R})=\phi(\mathbf{x}) \chi(\mathbf{R})$.
However, some degree of separability can be examined via the Schmidt decomposition. For example, in the case of a BO WF in Eq.~\eqref{eq:Born-Oppenheimer_wave_function}, the position matrix representation of the reduced density matrices reads 
\begin{widetext}
\begin{align}
        \left< \mathbf{R}' \left| \hat{\rho}^{\mathbf{R}} \right| \mathbf{R}'' \right> &=  \chi_{n,m} \left( \mathbf{R}' \right) \chi_{n,m}^*\left( \mathbf{R}'' \right ) \int{ d \mathbf{x} \phi_n \left( \mathbf{x};  \mathbf{R}' \right) \phi_n^* \left( \mathbf{x};  \mathbf{R}'' \right) },
        \label{eq:density_operator_R_general} \\
        \left< \mathbf{x}' \left| \hat{\rho}^{\mathbf{x}} \right| \mathbf{x}'' \right> &=  \int{ d \mathbf{R} \phi_n \left( \mathbf{x}';  \mathbf{R} \right) \phi_n^* \left( \mathbf{x}'';  \mathbf{R} \right) \chi_{n,m}\left( \mathbf{R} \right )\chi_{n,m}^* \left( \mathbf{R} \right) }.
        \label{eq:density_operator_x_general}
\end{align}
\end{widetext}
Instead, for a BH wave function, Eq.~\eqref{eq:Born-Huang_wave_function}, the reduced density matrix elements now are
\begin{widetext}
\begin{align}
        \left< \mathbf{R}' \left| \hat{\rho}^{\mathbf{R}} \right| \mathbf{R}'' \right> &= \sum_{n,n'} \sum_{m,m'} \Lambda_{nm} \Lambda_{n',m'}\chi_{n,m}(\mathbf{R}')\chi_{n',m'}^*( \mathbf{R}'')\int d \mathbf{x} \phi_{n}(\mathbf{x} ; \mathbf{R}' ) \phi_{n'}^*( \mathbf{x};  \mathbf{R}''),
        \label{eq:density_operator_R_BH} \\
        \left< \mathbf{x}' \left| \hat{\rho}^{\mathbf{x}} \right| \mathbf{x}'' \right> &=  \sum_{n,n'} \sum_{m,m'}\Lambda_{n,m} \Lambda_{n',m'} \int d \mathbf{R} \phi_n( \mathbf{x}';  \mathbf{R}) \phi_{n'}^*( \mathbf{x}'';  \mathbf{R}) \chi_{n,m}( \mathbf{R})\chi_{n',m'}^*( \mathbf{R}).
        \label{eq:density_operator_x_BH}.
\end{align}
\end{widetext}
and an analogous form for the variational total wave function.
The Schmidt WF, obtained after diagonalizing the reduced nuclear and electronic density operators, is now represented as a sum of products of Schmidt eigenfunctions
\begin{equation}
\label{eq:SchmidtWF}
    \Psi(\mathbf{x},\mathbf{R}) = \sum_i^{\mathrm{min} (M,N)} \sqrt{\lambda_i} 
    u_i(\mathbf{x})  v_i(\mathbf{R}).
\end{equation}
This expansion (which works for the BO wave function, the BH expansion and the variational ansatz) always contains some degree of entanglement between electronic and nuclear halfspaces and reaches a maximum entanglement entropy for $\sqrt{\lambda_i} = 1/\sqrt{\mathrm{min}(M,N)}$, according to Eq.~\eqref{eq:entropy}.

\subsection{\label{sec:discrete_ham}Discrete representation of the molecular Hamiltonian}

To implement numerically the method of solution for both the electronic and the nuclear motions, several methods based on finite grids are available. We choose the Fourier grid Hamiltonian (FGH) method \cite{Marston1989}, which belongs to the family of discrete variable representation (DVR) approaches \cite{Light1985}. In the FGH method, a continuous spatial variable $r$ is discretized as
\begin{equation}
    r_i = i \Delta r,   
    \label{eq:FGH_grid}
\end{equation}
with spacing $\Delta r$ for a total number of points $N$. The kinetic energy is represented in momentum space with a spacing given by $\Delta k = 2\pi / (N \Delta r)$. The latter spacing in $k$-space defines the longest wave length (smallest frequency) that can be resolved.
The Hamiltonian matrix for a single degree of freedom on this grid is then constructed as
\begin{equation}
     H_{ij} = 
    \frac{2}{N} \sum_{n=1}^{(N-1)/2} \cos\left( 2\pi n\frac{(i-j)}{N} \right) T_n + V(r_i) \delta_{ij},
    \label{eq:FGH}
\end{equation}
where $T_n = \frac{1}{2m} (n \Delta k)^2$ is the discretized kinetic energy term, and $V(r_i)$ is the potential energy at each grid point. For a full derivation, see Ref.~\cite{Marston1989}. The eigenvectors resulting from this FGH are composed directly of the WF amplitudes at each grid point. Then $\psi =  \sum_{i=1}^N \psi(r_i)$ and the norm $\int d r \left| \psi (r) \right|^2 =1$ in the FGH method is given by $ \sum_{i=1}^{N} \left| \psi(r_i) \right|^2 = 1$.

\subsubsection{\label{sec:FGH_BO} \textbf{FGH for a molecular problem in one dimension}}

In this work (we believe without much loss of generality) we solve the
electro-nuclear molecular problem for 1D systems in which both electrons and nuclei move in the same axis. 
This means that we neglect the rotation for the nuclear problem, 
and we propose simplified BO WFs in the form $\phi_n(x;R)\chi_{m,n}(R)$, where $R$ is just the vibrational coordinate.   

\paragraph{Electronic Hamiltonian on a Grid.}
The electronic Hamiltonian [Eq.~\eqref{eq:electronic_hamiltonian}] (now in 1D) is solved for each fixed internuclear distance $R$, thus 
diagonalizing the FGH matrix, Eq.~\eqref{eq:FGH}, in the electronic grid $\{x_j\}$. Note that the potential energy operator $\hat{V}$ depends parametrically on $R$. We are then provided with a set of electronic eigenfunctions $\phi_n(x_j;R)$, along with the corresponding eigenvalues $E_n(R)$, which become the potential energy curves (PEC) for the 1D vibrational problem.

\paragraph{Nuclear Hamiltonian on a Grid.}
The nuclear motion is solved with the nuclear Hamiltonian in Eq.~\eqref{eq:eq:nuclear_hamiltonian} for 1D. We provide a grid also for the internuclear distances 
$\{R_j\}$ and the nuclear FGH Hamiltonian is diagonalized for the PEC of each electronic state $n$. Then we have at our disposal a set of vibrational eigenfunctions $\chi_{n,m}(R_j)$ represented on the nuclear grid, along with their corresponding total vibronic energies $W_m^n$.
\subsubsection{\textbf{Reduced density matrices in the FGH discrete basis}}

In the FGH representation, the matrix elements for the electronic and nuclear reduced density matrices are now discretized within the 1D grids $\{x_k\}$ and $\{R_i\}$. For instance, for the case of BO WFs they are
\begin{widetext}
\begin{align}
        \left< R_i \left| \hat{\rho}^{R} \right| R_j \right> &=  \chi_{n,m} \left( R_i \right) \chi_{n,m}^*\left( R_j \right ) \sum_{k} {  \phi_n \left( x_k;  R_i \right) \phi_n^* \left( x_k;  R_j \right) },
        \label{eq:density_operator_R} \\
        \left< x_i \left| \hat{\rho}^{x} \right| x_j \right> &=  \sum_{k}{ \phi_n \left( x_i;  R_k \right) \phi_n^* \left( x_j;  R_k \right) \chi_{n,m}\left( R_k \right )\chi_{n,m}^* \left( R_k \right) },
        \label{eq:density_operator_x}
\end{align}
\end{widetext}
and similarly for the Born-Huang expansion and the total variational ansatz. Here we choose for the ongoing analysis the nuclear reduced density matrix in Eq.~\eqref{eq:density_operator_R}, which allows for a clearer interpretation.

\begin{figure}[t!]
    \centering\includegraphics[scale=0.41]{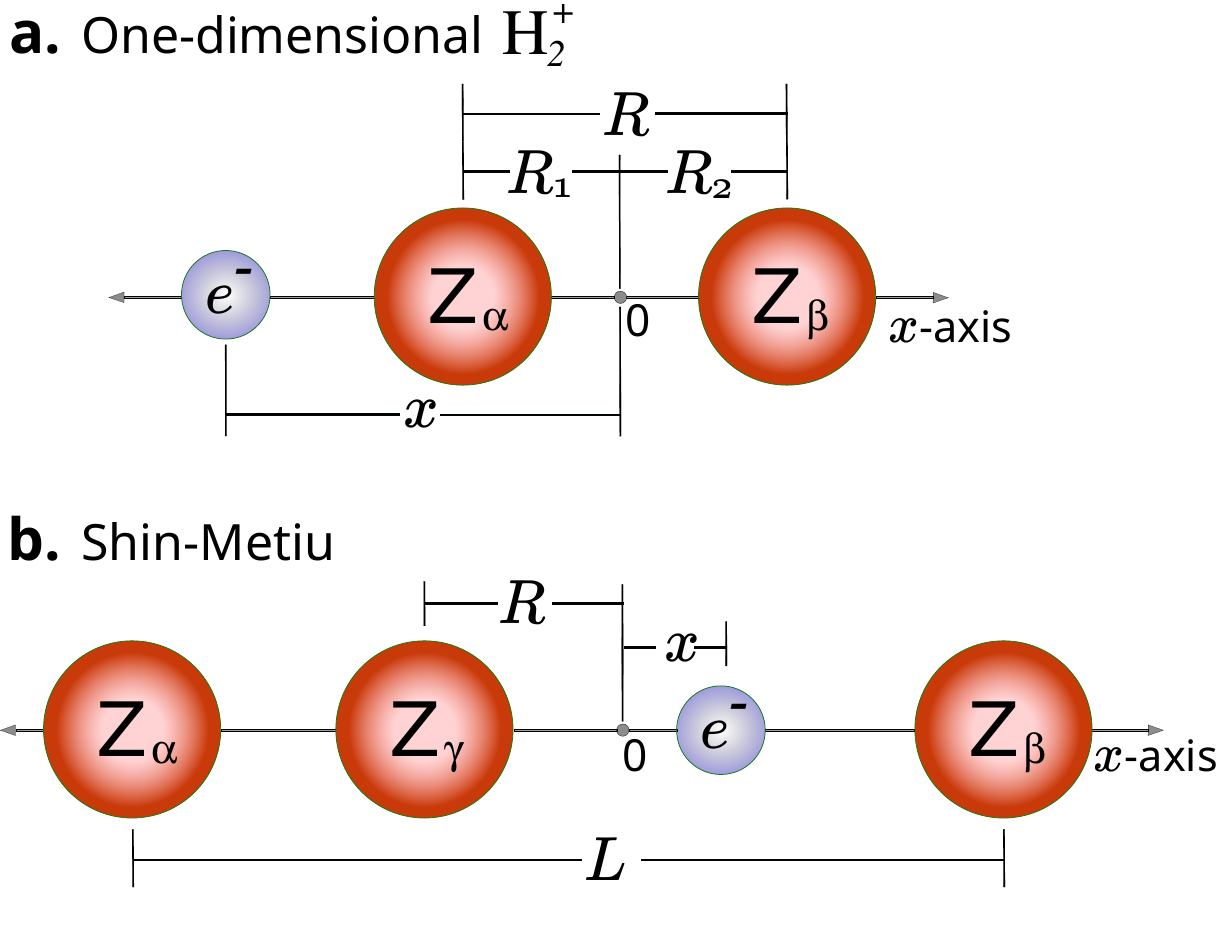}
    \caption{ (a) Schematic representation of the one-dimensional H$_2^+$ molecule, with motions constrained along the $x$-axis. The nuclei are located at positions $R_1$ and $R_2$ with respect to the nuclear center of mass, and $R$ denotes the internuclear distance. The variable $x$ represents the electronic degree of freedom relative to the nuclear center of mass. (b) Diagram of the Shin-Metiu model. Two ions with nuclear charges $Z_\alpha$ and $Z_\beta$ are fixed in space and separated by a distance $L$, while a third ion with charge $Z_\gamma$ moves freely. The electron also moves along the $x$-axis. For simplicity, all nuclear charges ($Z_\alpha$, $Z_\beta$, and $Z_\gamma$) in both models are set to unity.
    }
    \label{fig:systems}
\end{figure}
 
\section{Results}
\label{sec:results}

Simplified models can provide valuable insight into the behavior of more complex molecular systems. A common approach involves reducing the dimensionality of the real system by focusing on the most relevant or active reaction coordinates. In this work, we study two illustrative low-dimensional models; the first is the hydrogen molecular ion in 1D and the second is the Shin-Metiu Hamiltonian model \cite{Shin1996}. Both systems must be solved following the guidelines in Section~\ref{sec:theory}, and support solutions in BO form $\Psi^{\text{BO}}(x,R)$ and in BH form $\Psi^{\text{BH}}(x,R)$ in Eqs.~\eqref{eq:Born-Oppenheimer_wave_function} and \eqref{eq:Born-Huang_wave_function}, respectively. Electronic states in H$^+_2$ hardly show avoided crossings within the manifold of the lowest excited states of the same symmetry (Wigner-von Neumann non-crossing rule). However, it is worth noting that a series of extremely broad avoided crossings in H$^+_2$ among particular orbitals passed unnoticed for many years \cite{rost1989}, but this will be of no concern here, since we limit ourselves to the two lowest electronic states for simplicity.

\subsection{ One-dimensional H$_2^+$}
\label{sec:H2}

Our first system considers the simplest molecule H$_2^+$ in reduced dimensionality. Although the 3D case admits an exact solution in confocal elliptic coordinates \cite{Power1973}, the exact analytical solution for 1D has not yet been worked out to our knowledge. However, this simplified model has been widely used for decades, for example, in the study of nonlinear multiphoton processes in molecules \cite{Kulander1996}. Also, the plots and analysis of the electronic and nuclear WFs in 1D are much more simplified than in higher dimensions.

In this reduced model, only two coordinates, the internuclear separation $R$ and the electronic coordinate $x$, measured with respect to the center of mass of the nuclei, are taken into account along a single axis [see Fig.~\ref{fig:systems}(a)].
\begin{figure}[b!]
    \centering\includegraphics[scale=0.395]{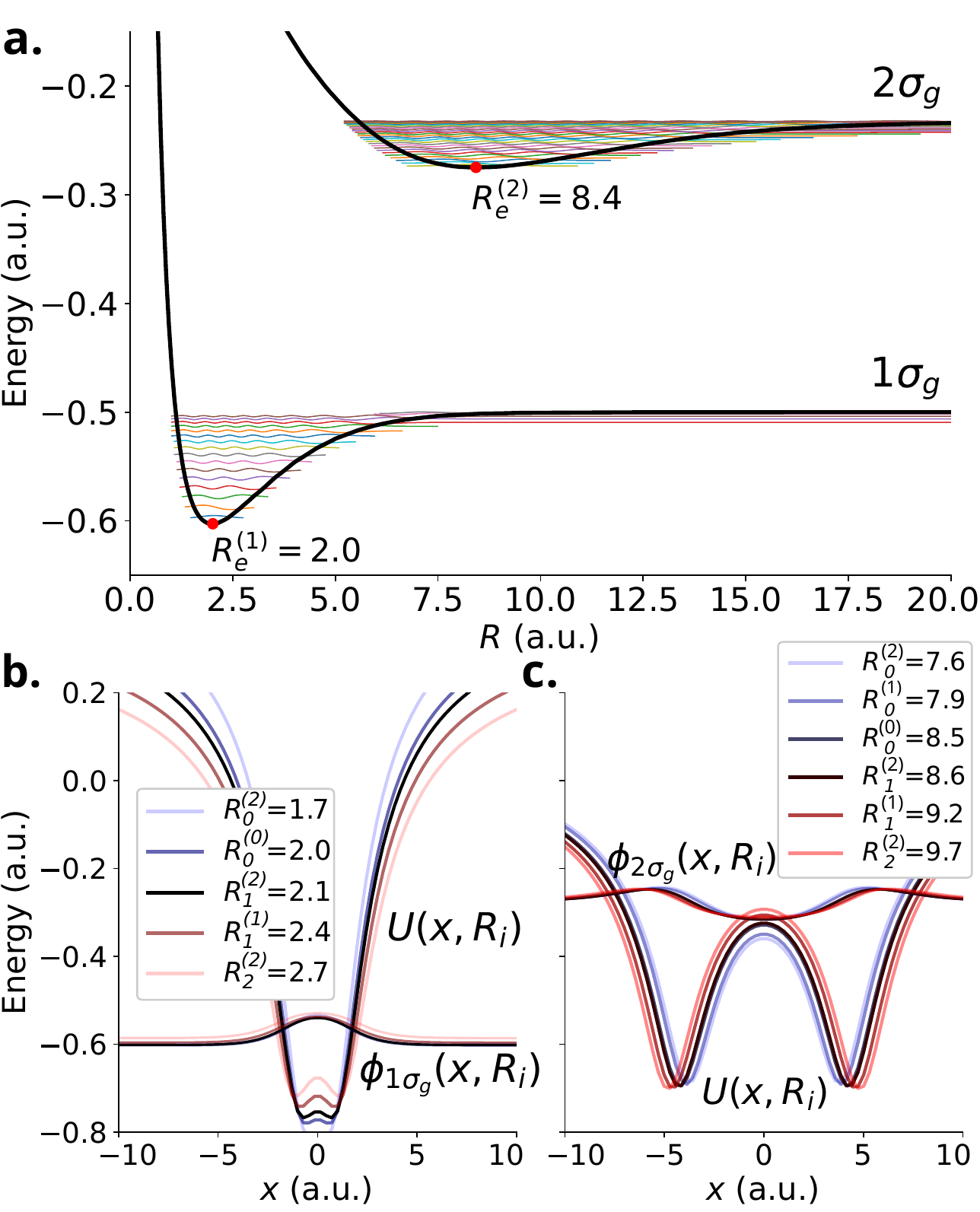}    
    \caption{(a) Vibrational bound states shifted to their corresponding variational energy associated to the BO potential energy curves for the molecular orbitals $1\sigma_g$ and $2\sigma_g$.
    (b)-(c) Electron-nuclei and nuclei-nuclei Coulomb electrostatic potential $\hat{U}(x, R_i)$ and the corresponding electronic wave functions for the $1\sigma_g$ and $2\sigma_g$ molecular orbitals, evaluated at selected nuclear configurations $R_i$ near the equilibrium bond lengths $R_e^{(1)} = 2.0$ a.u. and $R_e^{(2)} = 8.4$ a.u., respectively.
    The chosen nuclear positions $R^{(m)}_i$ correspond to the coordinates where the vibrational wave functions with labels $m$=0, 1 and 2 show their first local maxima and minima labeled with $i$=0, 1 and 2 (see also Fig.~\ref{fig:vibrational_states}). }
    \label{fig:POC_H2+}
\end{figure}
In the associated Hamiltonian, 
\begin{align}
    \displaystyle 
    \hat{H} & = \frac{\hat{P}_R^2}{M_P} +
            \frac{\hat{p}_x^2}{2\mu}  + 
            \frac{Z_\alpha Z_\beta }{R}  
           \nonumber\\
           & -\frac{Z_\alpha}{\sqrt{( \hat{x}+ \frac{R}{2} )^2 + a ( R )}} -
            \frac{Z_\beta}{\sqrt{ ( \hat{x}- \frac{R}{2} )^2 + a ( R ) }}, 
    \label{eq:hamiltoniano_H2+}
\end{align}
$\mu$ is the reduced electron mass and $M_p$  is the proton mass. A softening parameter $a( R )$ is used for the Coulomb singularity. It is chosen to produce the exact three-dimensional $1s\sigma_g$ BO PEC \cite{Yue2014,Madsen2012}. In this bipartite system, with three particles (nuclei with charges $Z_\alpha$ and $Z_\beta$ and one electron), the vibration and the electronic motion span the two Hilbert spaces $\mathcal{H}_R$ and $\mathcal{H}_x$. 

The solution of the Schr\"odinger equation associated with the Hamiltonian in Eq.~\eqref{eq:hamiltoniano_H2+}, follows the method described in Section~\ref{sec:FGH_BO}. We obtain electronic WFs in different nuclear configurations $R_i$ for the bonding orbitals $\phi_{1\sigma_g} (x;R)$ and $\phi_{2\sigma_g} (x;R)$, along with their associated vibrational states $\chi_{1\sigma_g/2\sigma_g v} (R)$, as illustrated in Fig.~\ref{fig:POC_H2+}. The PEC for the electronic state $1\sigma_g$ has 18 vibrational bound states up to the dissociation limit at $E = -0.5$ a.u. while the PEC for the state $2\sigma_g$ supports 28 vibrational bound states up to $E = -0.23$ a.u. (note that $a(R)$ is only adjusted to produce the exact dissociation limit of the ground state).

\begin{figure}[t]
    \centering\includegraphics[scale=0.39]{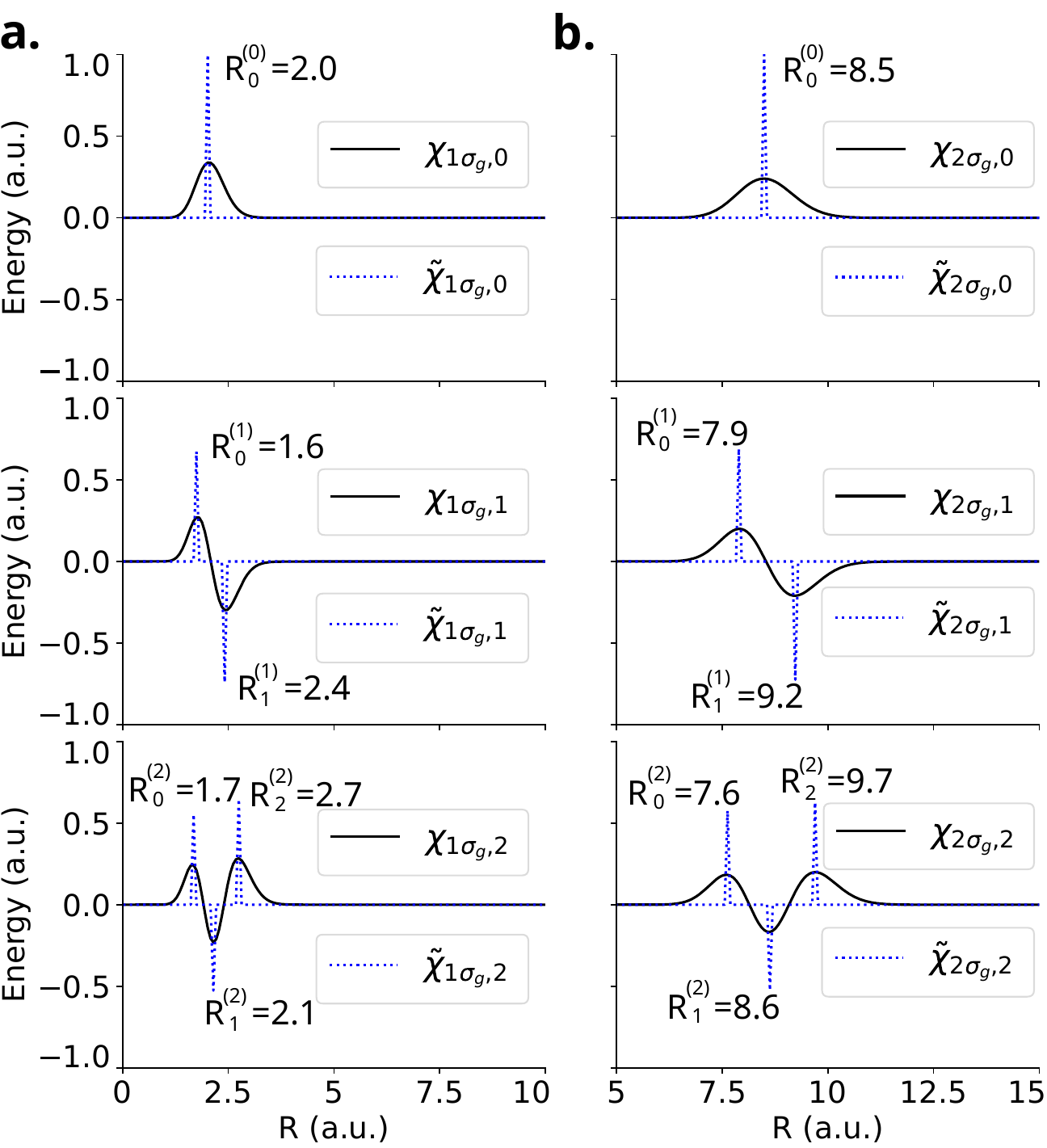}
    \caption{Vibrational wave functions for the one-dimensional H$_2^+$ molecular model. The numerical vibrational states $\chi_{1\sigma_g/2\sigma_g,m}$ are shown in black solid lines, while their corresponding simplified versions $\tilde{\chi}_{1\sigma_g/2\sigma_g,m}$, constructed using Eq.~\eqref{eq:simplified_vibrationa_wave_function}, are shown in blue for $m=$0, 1 and 2. The marked positions $R^{(m)}_i$ indicate the internuclear distance at which the vibrational wave function with label $m$ shows maxima or minima, with the ordering labels $i=0,1,2...$; (a) Lowest three vibrational wave functions of the electronic ground state $1\sigma_g$. (b) Lowest three vibrational wave functions of the electronic excited state $2\sigma_g$. Both 
    $\chi_{1\sigma_g/2\sigma_g,m}$ and $\tilde{\chi}_{1\sigma_g/2\sigma_g,m}$ states are normalized to unity.  
    }
    \label{fig:vibrational_states}
\end{figure}

\begin{figure}[t!]
\centering\includegraphics[scale=0.375]{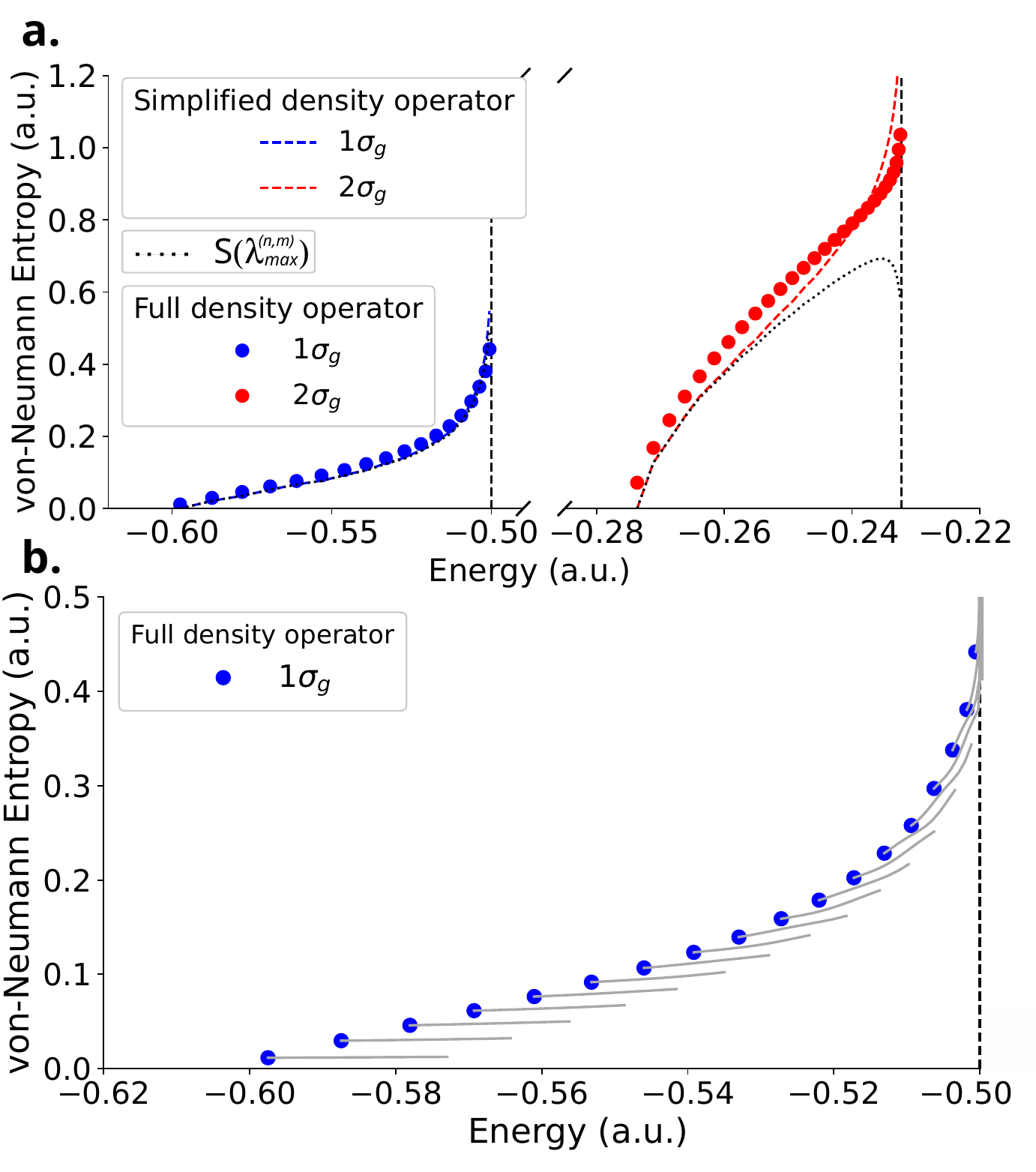}
    \caption{(a) Von Neumann entropy as a function of the energy of the vibronic state for the $1\sigma_g$ PEC (blue dots) and for the $2\sigma_g$ PEC (red dots). Each plot includes the corresponding bound vibrational states, i.e. the states below the black dashed vertical lines for each case. The dashed blue and red lines correspond to the von Neumann entropies computed using the simplified model of the density matrix explained in Section~\ref{sec:simplified_density} and the black doted lines denote the von Neumann entropies computed using the approach described by Eq.~\eqref{eq:entropy_approach}. (b) von Neumann entropy in 3D H$^+_2$ of rovibrational states as a function of the energy $W^J_{1s\sigma_g,m}$ (blue dots are the same as in Fig.~\ref{fig:entropy_H2+}(a) for 1D). Solid grey lines connect the entropy pertaining to the same vibrational number $m$, for rotational states from $J=0$ up to $J=15$.
    }
    \label{fig:entropy_H2+}
\end{figure}

For each electro-nuclear BO state, we have computed the entanglement content through the von Neumann entropy, either using Eq.~\eqref{eq:density_operator_R} or Eq.~\eqref{eq:density_operator_x} for the discrete FGH representation of the reduced density matrices and Eq.~\eqref{eq:entropy}, with the obtained set of eigenvalues $\{ \lambda_i \}$. 
The high amount of oscillations in the vibrational grid makes that in practice the number of points in the FGH representation of the nuclear grid be larger than the one in the electronic grid ($M>N$). 
In Fig.~\ref{fig:entropy_H2+}(a), we include the von Neumann entropy associated with the vibronic bound states of the electronic states $1\sigma_g$ and $2\sigma_g$. 
Both cases exhibit a monotonic increase in the amount of entanglement $S$ with the vibrational excitation $v$ and satisfy $S_{1\sigma_g,v} < S_{2\sigma_g,v}$. 
The latter effect is in principle related to the spatial extension of the vibrational WF. 
In general, the more spreading of the WF in the configuration space, the more delocalization and then its related entropy increases. In fact, above the dissociation threshold the vibronic WFs pertain to the nuclear continuum (states with infinite extension) and the entanglement must diverge for the exact continuum states.
The entanglement for the electronic state $1\sigma_g$ has a convex curvature, whereas that for $2\sigma_g$ has a concave curvature. 
A more detailed explanation is given below using a simplified approach for the reduced densities in Eq.~\eqref{eq:density_operator_x}.

\paragraph{Contribution of molecular rotation to entanglement.} 

At this point, we may wonder how molecular rotation may affect the electro-vibrational entanglement. 
For simplicity, we may assume a 3D nuclear WF in the form $\chi^{J,M}_{n,m} ({\bf R}) = \chi^J_{n,m}(R) \mathcal{Y}^M_J (\theta,\phi)$, where $\mathcal{Y}^M_J (\theta,\phi)$ is a spherical harmonic function.
For a rotating diatomic molecule, the reduced nuclear Schrödinger equation for the radial part $\chi^J_{n,m}(R)$ reads
\begin{equation}
\left[ -\frac{1}{2 \mu_N} \frac{\partial^2}{\partial R^2}
+ \frac{J(J+1)}{2 \mu_N R^2}
 + E_{n}(R) - W_{J,m}^{n} \right]
 \chi^J_{n,m}=0,
 \label{eq:angular_schrodinger}
\end{equation}
where $\mu_N$ denotes the nuclear reduced mass.
The entanglement for the resulting rovibrational states is shown in Fig.~\ref{fig:entropy_H2+}(b), for rotational quantum numbers from $J=0$ to $J=15$. 
The blue dots represent all bound states of the $1\sigma_g$ PEC in the 1D system. The horizontal gray line connects states with the same vibration number $m$ but different rotational number $J$. 
Although there are shifts in energy induced by the additional centrifugal potential, the von Neumann entropy remains nearly unaffected for the lowest vibrational levels. 
In contrast, for high vibrational excitation, the entanglement contributed by rotation within the same vibrational level seems to also follow a convex pattern with increasing rotational excitation. In conclusion, our 1D model for H$^+_2$ already captures the leading vibrational part of the entanglement.

\subsubsection{\label{sec:simplified_density}\textbf{Model with a simplified density matrix}}

We begin by simplifying the vibrational WF $\chi_{n,m} (R)$, defined on a discrete vibrational grid with points $\{ R_i \}_{i=1}^M \in {\mathbf F}$, as obtained via the FGH method. All vibrational WFs exhibit oscillatory behavior, and we propose to reduce the representation of the FGH
with a subset ${\mathbf S} \subset {\mathbf F}$ of grid points that correspond only to the location of maximum amplitudes in the WF $\{ R_k \}_{k=0}^{m} \in {\mathbf S}$. The new amplitudes $\tilde{\chi}_{n,m} (R_k)$ at each point $R_k$ are conveniently scaled to produce a new normalized state. The number of extrema (maxima and minima) in a vibrational wave function $\chi_{n,m} (R)$ with $m$ nodes is determined by the vibrational quantum number $m$, therefore with $m+1$ grid points in the new reduced set. Consequently, the simplified (normalized) vibrational WF $\tilde{\chi}_{n,m}(R)$ at the grid point $R_k$ reads
\begin{equation}
    \tilde{\chi}_{n,m}(R_k) = \chi_{n,m}(R_k) / \sqrt{\sum_{\ell=0}^m \chi_{n,m}(R_\ell) }.
    \label{eq:simplified_vibrationa_wave_function}
\end{equation}
Within the FGH grid, this construction ensures that the trace of the density matrix remains equal to unity, due to the renormalization of the quantum states.
Fig.~\ref{fig:vibrational_states} shows the lowest three vibrational WFs ($\chi_{1\sigma_g, m}$ and
$\chi_{2\sigma_g, m}$ for $m=0,1,2$) for the 1D H$^+_2$ molecular system, along with their counterparts in our simplified model
($\tilde{\chi}_{1\sigma_g, m}$ and
$\tilde{\chi}_{2\sigma_g, m}$ for $m=0,1,2$). Using
Eq.~\eqref{eq:simplified_vibrationa_wave_function}, we obtain a much simplified expression for the density matrix, given by
\begin{equation}
    \tilde{\rho}_{k,\ell}^{(R)} = \tilde{\chi}_{n,m}(R_k) \tilde{\chi}_{n,m}(R_\ell) S_{n}(R_k,R_\ell),
    \label{eq:simplified_density_R}
\end{equation}
where $S_{n}(R_k,R_\ell)$  denotes the overlap integral between electronic WFs at nuclear distances $R_k$ and $R_\ell$, i.e. $S_{n}(R_k,R_\ell) =  \int{ d x \phi_n \left( x;  R_k \right) \phi_n^* \left( x;  R_\ell \right) }$. In this discrete representation, the dimension of the simplified density matrix corresponds to the number of extrema in the vibrational state. 
For example, the vibrational state $\tilde{\chi}_{n,0}$ leads to a density matrix of dimension unity, with zero entanglement by construction, while the state $\tilde{\chi}_{n,17}$ results in a dimension $18\times18$ for the matrix.

The overlap integrals, $0 \le S_{n}(R_k,R_\ell) \le 1$, thus introduce weights in the elements of the reduced density matrix. In the limit of perfect overlap, when $S_{n}(R_k,R_\ell)=1$, for all $R_k$ and $R_\ell$, it implies that the electronic WF is the same for
all internuclear distances and the BO state is then fully separable and not entangled between electronic and nuclear motions. This is consistent with the fact that the reduced density matrix $\tilde{\rho}_{k,\ell}^{(R)} = \tilde{\chi}_{n,m}(R_k) \tilde{\chi}_{n,m}(R_\ell)$ and 
its eigenvalues
are determined solely by the vibrational functions.
This density matrix is a rank-one matrix, which means that the matrix is generated by the outer product of a vector with itself, ${\mathbf M}={\mathbf x} {\mathbf x}^T$, and has at most one non-zero eigenvalue $\lambda={\mathbf x}^T{\mathbf x}$ equal to the trace. Since vibrational states are normalized within the FGH grid, then $\text{Tr} \left[ \tilde{\rho}^{(R)} \right ] = 1$. Consequently, we also have $\text{Tr} \left[ \left( \tilde{\rho}^{(R)} \right)^2 \right] = 1$, the von Neumann entropy vanishes, and the fully separable BO state is not entangled. However, any molecule departs from this ideal situation.

Fig.~\ref{fig:entropy_H2+} shows that the von Neumann entropy computed using the simplified density matrix closely matches the results obtained from the full density matrix; specifically, it also satisfies the asymptotic behavior at the dissociation thresholds. The subset with pairs of maxima and minima $\{R_k, R_\ell \} \in {\mathbf S}$ in the vibrational WFs suffices to monitor the variation of the electronic WFs in those different nuclear geometries. 
For example, the lowest three vibrational WFs $\chi_{1\sigma_g,n} (R)$ are mainly localized in the interval $1.7 < R < 2.7$, for which the electronic WFs barely change, as shown in Fig.~\ref{fig:vibrational_states}(a)-(b) (the electronic overlaps at the grid points $\{R_k, R_\ell \}$ remain close to unity). 

For higher vibrational excitations, the WFs increase their spatial extension, and their extrema (or their nodes) are more separated from each other. 
Electronic overlaps noticeably depart from unity for many distant pairs $\{ R_k, R_\ell \}$ and these overlaps reduce the magnitude of the off diagonal density matrix elements. 
This effect is even more pronounced in the electronic excited state $2\sigma_g$ for which the distributions of pairs $\{R_k, R_\ell \}$ are even more separated than in $1s\sigma_g$ and the overlap factor assesses very different electronic densities at two distant points. 

In this way, nuclear WFs act as testers of the similarity of the electronic WFs at different geometries. The most relevant matrix elements contributing to the eigenvalue spectrum of the reduced density matrix are those that pertain to bands close to the diagonal; for a fixed row $k$, close to the diagonal, the electronic wave function is monitored with $S_n(R_k, R_\ell)$ at adjacent nuclear points for $R_\ell$, $\{ R_{k-1}, R_k, R_{k+1}\} $. The worse the overlap in this tridiagonal matrix, the larger the entanglement since the eigenspectrum contains several sizeable eigenvalues $0 < \lambda_i < 1$.

\begin{figure*}[t]
\centering\includegraphics[scale=0.38]{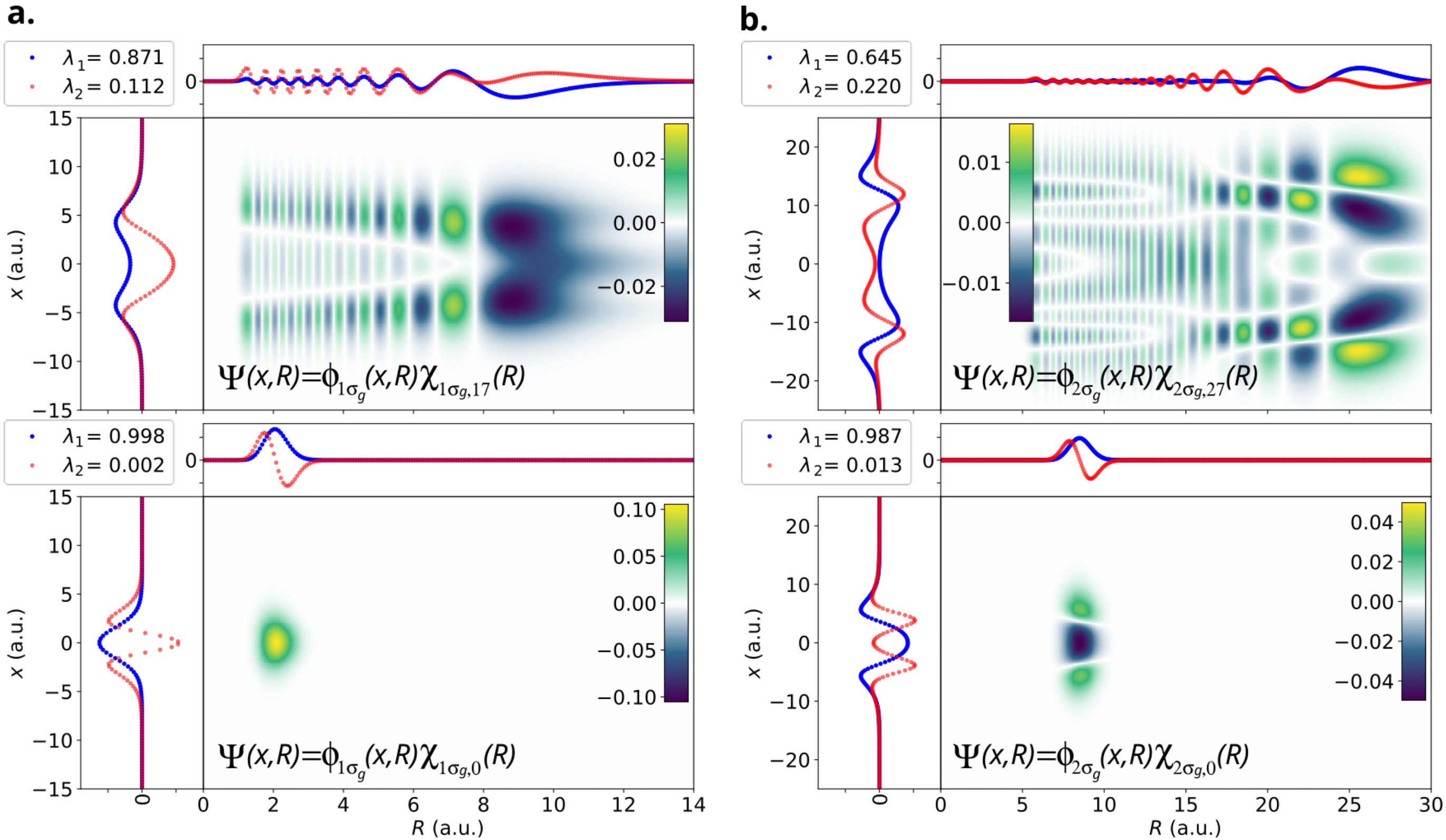}
    \caption{Total wave function under the Born-Oppenheimer approximation for the one-dimensional H$_2^+$ model. The central panels display the full electron-nuclear wave function, while the side panels present the first two Schmidt modes for both the electronic and nuclear subsystems, along with their corresponding largest Schmidt coefficients. (a) Lowest and highest vibrational eigenstates associated with the $1\sigma_g$ PEC. (b) Lowest and highest vibrational eigenstates associated with the $2\sigma_g$ PEC. The reconstructed wave functions obtained through the Schmidt decomposition accurately reproduce the original BO wave functions.
    }
\label{fig:schmidt_H2+}
\end{figure*}

Beyond the previous analysis, for large overlap values, 
$S_{n}(R_k,R_\ell) \sim 1$, we can introduce a perturbation to the reduced density rank-one matrix defined solely by the vibrational WFs. Then, the simplified density matrix from Eq.~\eqref{eq:simplified_density_R} can be recast as
\begin{equation}
    \tilde{\rho}_{k,\ell}^{(R)} =   \tilde{\chi}_{n,m}(R_k) \tilde{\chi}_{n,m}(R_\ell) + \epsilon_{k,\ell}^{(n,m)},
    \label{eq:perturbation_simplified_density_R}
\end{equation}
where $\epsilon_{l,\ell}^{(n,m)}$ is a perturbation defined as $\epsilon_{k,\ell}^{(n,m)}$ = $- \tilde{\chi}_{n,m}(R_k) \tilde{\chi}_{n,m}(R_\ell)\left[ 1 - S_n(R_k,R_l) \right]$. In the case of perfect overlap, the factor $1 - S_n(R_k,R_\ell)$ vanishes and $\tilde{\rho}^{(R)}$ reduces to the rank-one matrix. For non-overlapping electronic WFs, the largest eigenvalue of $\tilde{\rho}^{(R)}$ takes the form
\begin{equation}
    \lambda^{(n,m)}_{max} = 1 + 2 \sum_{k<\ell=0}^{m} \epsilon_{k,\ell}^{(n,m)}\; \tilde{\chi}_{n,m}(R_k) \tilde{\chi}_{n,m}(R_\ell) < 1.
\end{equation}
In such a case, for moderately entangled states, it is often sufficient to consider only the two largest eigenvalues of the reduced density matrix to provide a good estimation of the entanglement content. By just considering the two largest eigenvalues, the von Neumann entropy takes a simple form in comparison with the one listed in Eq.~\eqref{eq:entropy}, given by
\begin{equation}
   \!\!\! S(\lambda_{max}^{n,m}) = -\log \left[ {\lambda_{max}^{(n,m)}}^{\lambda_{max}^{(n,m)}} 
    \left( 1 - \lambda_{max}^{(n,m)} \right)^{1-\lambda_{max}^{(n,m)}}   \right].
    \label{eq:entropy_approach}
\end{equation}
This approximate von Neuman entanglement is included in Fig.~\ref{fig:entropy_H2+}(a) with black dotted lines. It works fine for entropies below 0.5 (for all bound vibrational states in $1s\sigma_g$ and up to the 8th vibrational state in curve $2\sigma_g$). From this perturbative approach, we conclude that even slight changes in the electronic overlap may induce (up/down) concavity in the entropy curves. 

Compact vibrational states bounded by narrow PEC are expected to display similar electronic overlaps in the region between the two turning points.
In this case, we find that the largest eigenvalue in the $2s\sigma_g$ state departs from 1 faster but still linearly with the vibrational excitation energy $E$, $\lambda_{max}^{(2\sigma_g,E)} \sim 1 - a E$ with a constant $a$. 
On the other hand, for compact vibrational states bounded by the narrow $1s\sigma_g$ PEC, we find a slower and nonlinear departure from unity with the vibrational excitation energy, which we can model with a nonlinear term $\lambda_{max}^{(1\sigma_g,E)} \sim 1 - a E^2$.
These simple trends for $\lambda_{max}^{(n,m)}$, introduced in Eq.~\eqref{eq:entropy_approach}, explain the different concavities of the von Neuman entropies with increasing energy, as shown in Fig.~\ref{fig:entropy_H2+}.\\ 

Fig.~\ref{fig:schmidt_H2+}(a) shows two BO wave functions for two vibrational states (the lowest $m=0$ and the last bound $m=17$) in the ground state $1\sigma_g$, $\phi_{1\sigma_g}(x,R)\chi_{1\sigma_g,0}(R)$ and $\phi_{1\sigma_g}(x,R)\chi_{1\sigma_g,17}(R)$, while Fig.~\ref{fig:schmidt_H2+}(b) shows two BO wave functions also for the lowest and the upper vibrational states, $\phi_{2\sigma_g}(x,R)\chi_{2\sigma_g,0}(R)$ and $\phi_{2\sigma_g}(x,R)\chi_{2\sigma_g,27}(R)$ (see also Fig.~\ref{fig:POC_H2+}). Here we use the whole set of grid points $\{R_i\}_{i=1}^M \in {\bf F}$.
However, the full BO wave function can also be exactly reproduced with the complete set of Schmidt basis in Eq.~\eqref{eq:schmidt_decomposition} (eigenstates of the reduced density matrices). 
For the vibrational state $v=0$ in state $1\sigma_g$, the BO wave function is nearly separable since the first Schmidt coefficient $\lambda_1 = 0.998$ dominates, and the product of the lowest Schmidt electronic and nuclear eigenfunctions, $u_1(x)$ and $v_1(R)$, are sufficient to accurately describe the quantum state. 
A similar situation occurs for the lowest vibrational state in state $2\sigma_g$ although $\lambda_1$ now departs more from unity. 
All this reflects an almost complete separability for the ground state, thus a very low electron-nuclear entanglement and an almost vanishing von Neumann entropy. It should be noted that separability is exact and the entanglement entropy is zero for the same BO state represented in the simplified grid $\{R_k\}_{k=0}^m \in {\bf S}$ described above, because the ground state is represented by only a single grid point. In practice, 
the BO wave function reproduced with the Schmidt basis (in the electronic and nuclear halfspaces) using only the two highest eigenvalues is 
visually indistinguishable from those included in Fig.~\ref{fig:schmidt_H2+}. 

\subsection{\label{sec:shin-metiu} The Shin-Metiu model}

In the previous section, we analyzed how changes in the configuration space affect the degree of entanglement in a molecule for which BO separability is appropriate. We now turn to examine how non-adiabatic effects, particularly those arising from avoided crossings between PECs, influence the entanglement content. The Shin-Metiu model has already been used in different scenarios, and it allows us to study the structure and dynamics through different kinds of PECs and avoided crossings, let them be sharp or broad.

The Shin-Metiu molecular model consists of three ions and a single electron in 1D [see Fig.~\ref{fig:systems}(b)]. The ions labeled $Z_\alpha$ and $Z_\beta$ are fixed at a distance of $L = 10\ \text{\AA} \approx 18.9$ a.u. 
The third ion, $Z_\gamma$, with nuclear mass $M$ is allowed to move along the $x$-axis between the two fixed ions, while the electron with mass $m_e$ moves freely along the same axis. 
The degrees of freedom are the position of the electron $x$, and the position $R$ of the ion with charge $Z_\gamma$ \cite{Seokmin1995,Shin1996}, with respect to the origin in $x=0$.
The Shin-Metiu Hamiltonian reads
\begin{align}
   & \hat{H} = \frac{ \hat{P}_R^2 }{ 2 M_P } +
            \frac{ \hat{p}_x^2 }{ 2m_e } +
            \frac{ Z_\alpha Z_\gamma }{ \left| \hat{R}+L/2 \right| } + 
            \frac{ Z_\beta Z_\gamma }{ \left| \hat{R}-L/2 \right| }  \nonumber\\
            &-\frac{ Z_\alpha \ \text{erf} \left(  \frac{\hat{x}}{R^{\alpha}_{c}} \right)} { \left| \hat{x}+L/2 \right| } -
            \frac{Z_\beta \ \text{erf} \left( \frac{\hat{x}}{R^{\beta}_{c}} \right) }{ \left|\hat{x}-L/2 \right| } -
            \frac{Z_\gamma \ \text{erf} \left( \frac{\hat{x}}{R^{\gamma}_{c}}\right) }{ |\hat{x}-\hat{R}| }.
    \label{eq:hamiltoniano_shin_metiu}
\end{align}
This model includes two Coulomb repulsion terms between the nuclei and three attractive electron-nucleus interactions. 
The latter are represented using screened Coulomb potentials in the form $V_{Ne} = Z_i \text{erf} \left( \hat{x}/R^{i}_{c} \right)/r_i$, where $r_i$ is the distance between the electron and the nucleus $i$, and $R_c^i$ is a cut-off parameter.

\begin{figure}
    \centering\includegraphics[scale=0.43]{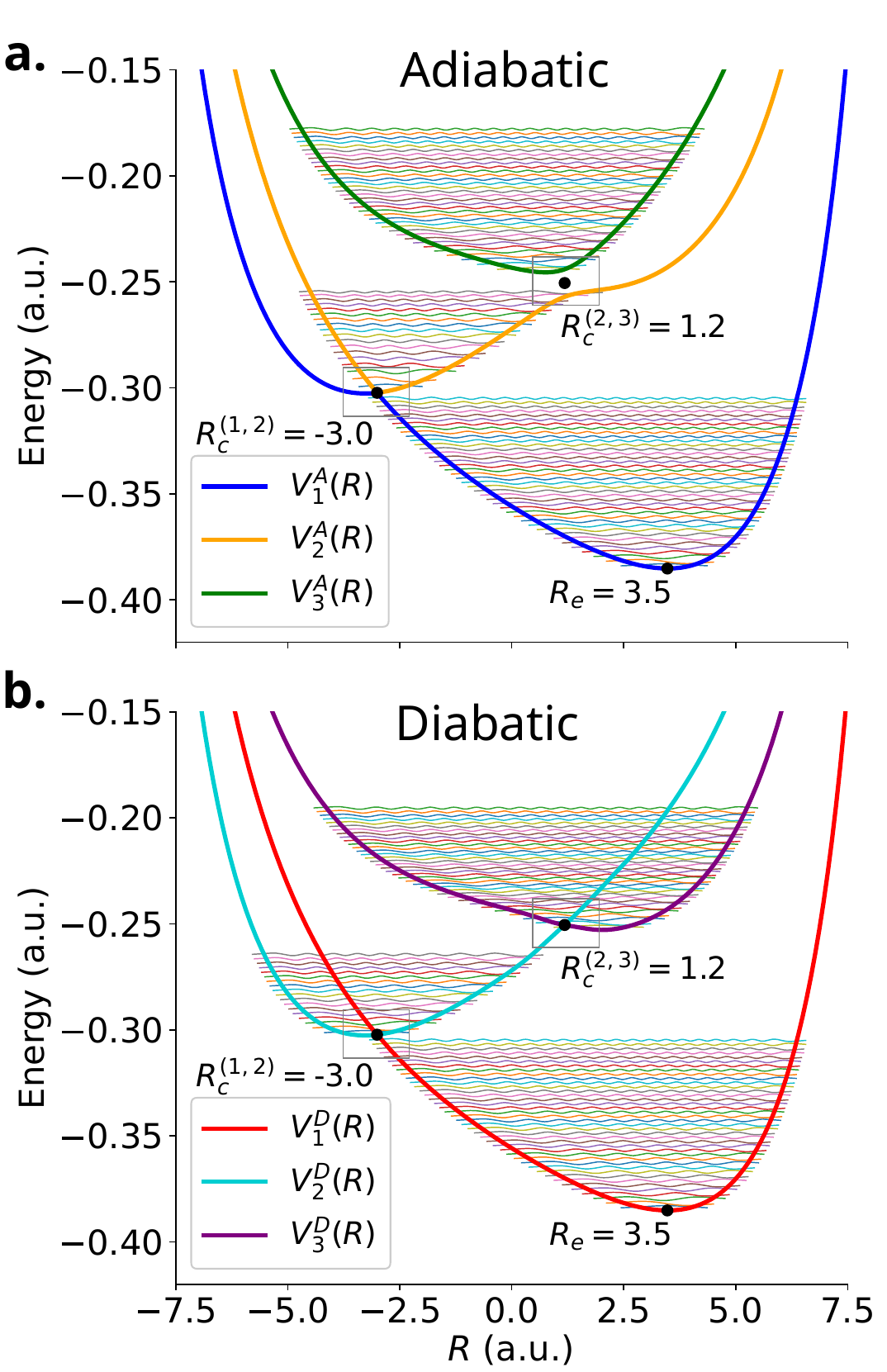}
    \caption{ (a) Adiabatic potential energy curves (PECs) for the Shin-Metiu model with screening parameters $R_c^\alpha = 3.00$ a.u., $R_c^\beta = 2.20$ a.u., and $R_c^\gamma = 4.00$ a.u., resulting in two avoided crossings at $R^{(1,2)} = -3.0$ a.u. and $R^{(2,3)} = 1.2$ a.u. (b) Diabatic PECs obtained by rotating the adiabatic states around each avoided crossing, as detailed in Appendix~\ref{sec:adiabatic-diabatic}.
    All vibrational eigenfunctions below the avoided crossings (adiabatic picture) or below the real crossings (diabatic picture) are included in the figure for each electronic state.
    }
    \label{fig:POC_shin-metiu}
\end{figure}

\begin{figure}[t]
    \centering\includegraphics[scale=0.395]{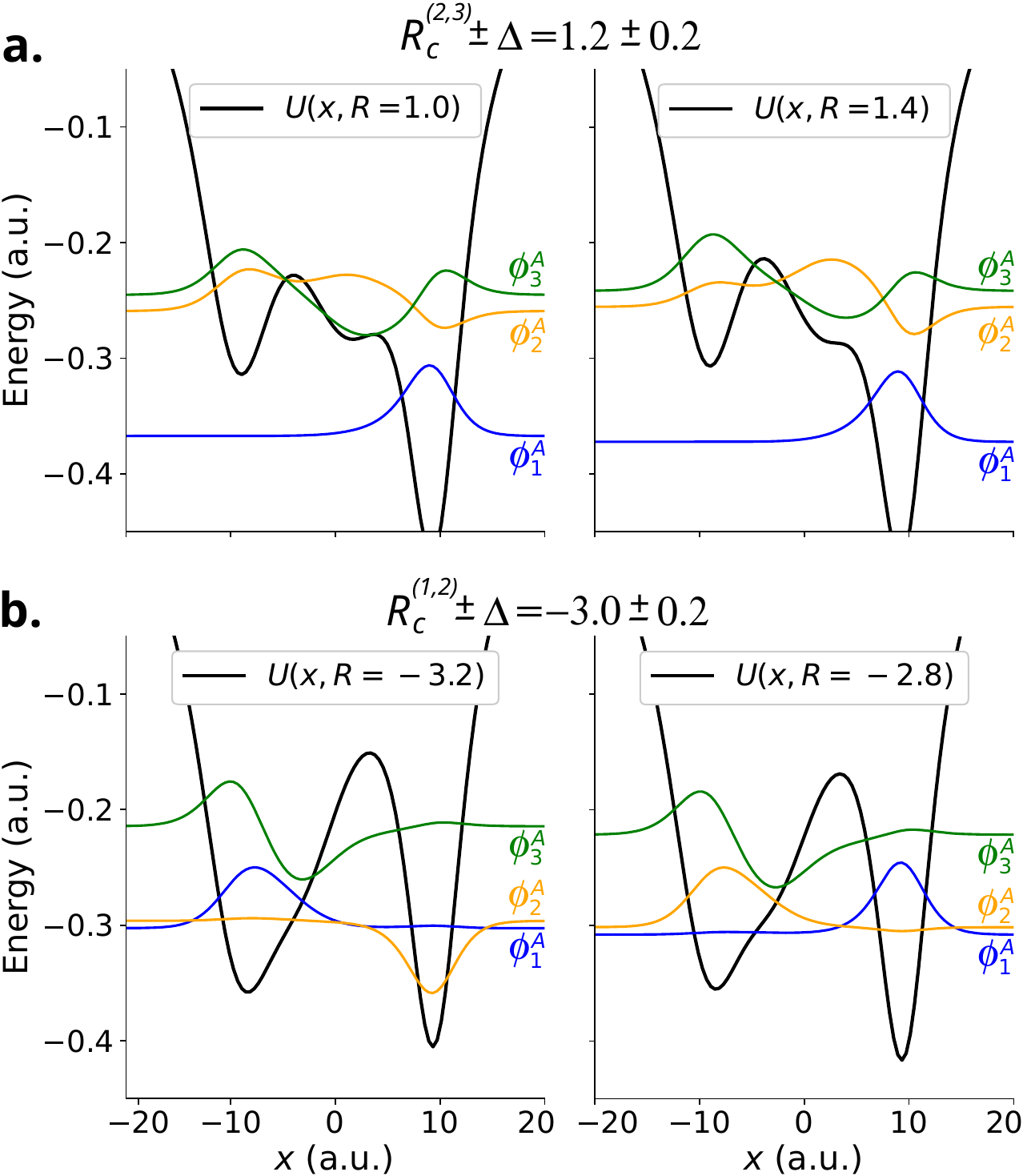}
    \caption{ {\bf Adiabatic picture.} BO electronic eigenfunctions 
    $\phi_1^A$ (blue), $\phi_2^A$ (yellow) and $\phi_3^A$ (green) of the Shin-Metiu Hamiltonian for nuclear distances $R$ shifted with $\Delta$ to the left and to the right from the locations of the avoided crossings $R_c^{(1,2)} = -3.0$ a.u. (panel a) and $R_c^{(2,3)} = 1.2$ a.u. (panel b) in Fig.~\ref{fig:POC_shin-metiu}(a). The potential $U(x,R)$ in Eq.~\eqref{eq:hamiltoniano_shin_metiu} is plotted with black solid line for different nuclear geometries $R$. 
    }
\label{fig:adiabatic_states}
\end{figure}
\begin{figure}[t]
    \centering\includegraphics[scale=0.395]{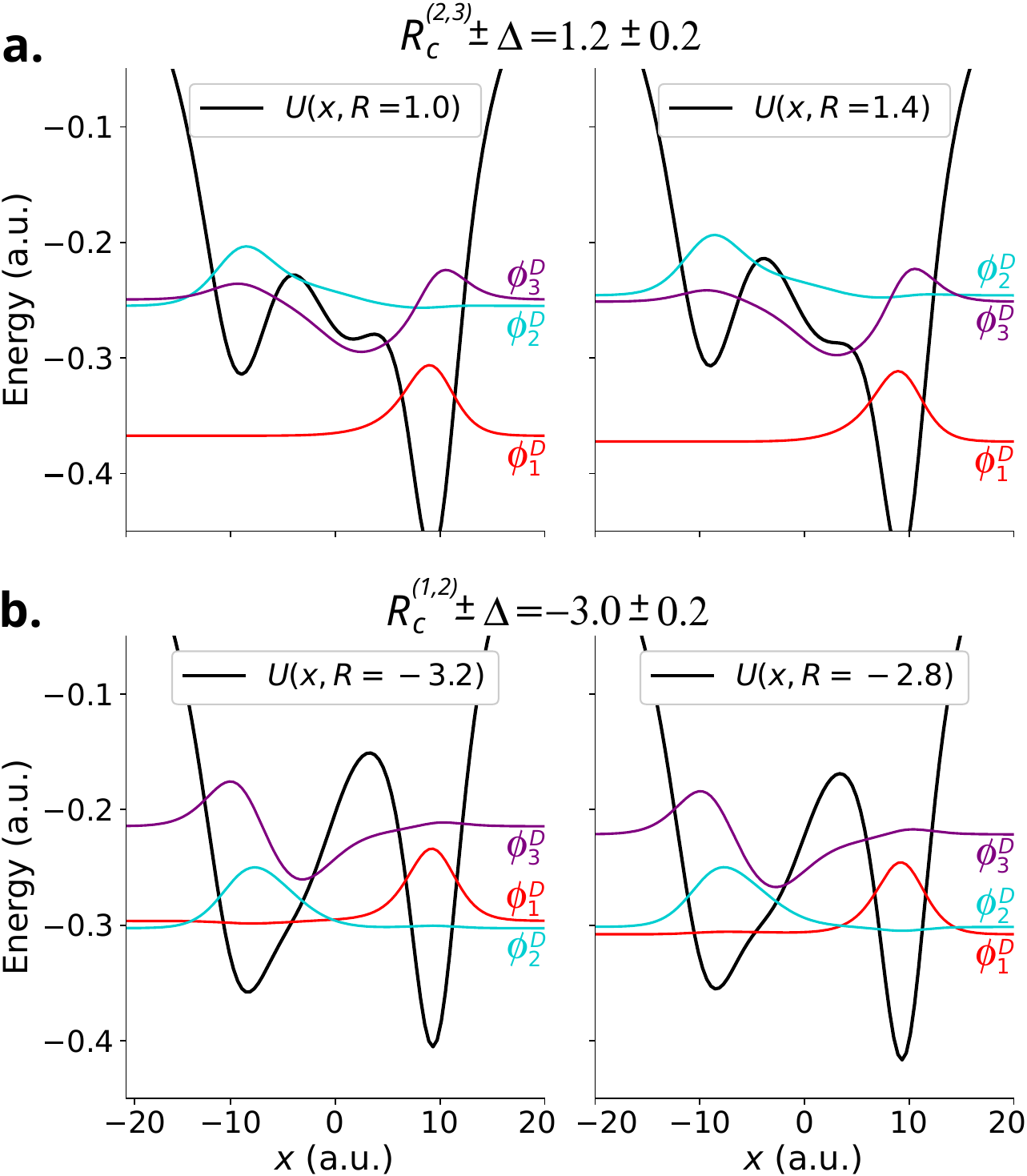}
    \caption{ {\bf Diabatic picture.} Electronic diabatic functions 
    $\phi_1^D$ (red), $\phi_2^D$ (cyan) and $\phi_3^D$ (purple) of the Shin-Metiu Hamiltonian for nuclear distances $R$ shifted with $\Delta$ to the left and to the right from the locations of the crossings $R_c^{(1,2)} = -3.0$ a.u. (panel a) and $R_c^{(2,3)} = 1.2$ a.u. (panel b) in Fig.~\ref{fig:POC_shin-metiu}(b). The potential $U(x,R)$ in Eq.~\eqref{eq:hamiltoniano_shin_metiu} is plotted with black solid line for the different nuclear geometries given by $R$. 
    }
\label{fig:diabatic_states}
\end{figure}

This model supports several stable nuclear configurations, corresponding to minima in the three adiabatic PECs. Adjusting charges and cut-off parameters $R_c^i$ allows us to adjust the electron’s ability to localize the ion with charge $Z_i$; different parameters can produce dramatic changes in adiabatic PECs and, in particular, in the strength and width of the avoided crossings (see \cite{Seokmin1995}). 
Throughout this work, we use a linear chain of protons $Z_\alpha = Z_\beta = Z_\gamma = 1$ and we select the cut-off parameters to be $R_c^\alpha = 3.00$ a.u. $R_c^\beta = 2.20$ a.u. and $R_c^\gamma = 4.00$ a.u. 
This choice produces the adiabatic PECs $V^A_1 (R)$, $V^A_2 (R)$ and $V^A_3 (R)$, shown in Fig.~\ref{fig:POC_shin-metiu}(a).

\subsubsection{\bf{Adiabatic and Diabatic states}}

In this setup, the non-adiabatic couplings between the three lowest electronic states are non-negligible. As shown in Fig.~\ref{fig:POC_shin-metiu}(a), the system exhibits two different avoided crossings: a sharp one between the curves $V_1^A$ and $V_2^A$ located at $R^{(1,2)} = -3.0$ a.u., and a broader one between the curves $V_2^A$ and $V_3^A$ at $R^{(2,3)} = 1.2$ a.u. 
At avoided crossings it is known that the BO approximation fails and the adiabatic states become coupled through non-adiabatic couplings. To force a BO approximation implies neglecting the non-adiabatic couplings. Strong non-adiabatic couplings cause character exchange between the states involved. For example, the electronic states
$\phi^A_1$ and $\phi^A_2$ fully exchange their character after the sharp avoided crossing, and it can be observed in Fig.~\ref{fig:adiabatic_states}(b) for the narrow avoided crossing located at $R_c=-3.0$ a.u. in Fig.~\ref{fig:POC_shin-metiu}(a). However, the broad avoided crossing at a distance $R_c=1.25$ a.u. (see Fig.~\ref{fig:POC_shin-metiu}(a)) between the curves $V_2^A$ and $V_3^A$ produces little character exchange after the avoided crossing, as illustrated in Fig.~\ref{fig:adiabatic_states}(a).
The presence of avoided crossings, non-adiabatic couplings and character exchanges indicates that the WF becomes a superposition of BO vibronic eigenstates, which ends up with a change in the electro-vibrational entanglement. 
For a briefing on the calculation of non-adiabatic couplings, see Appendix~\ref{non-adiabatic_couplings}.

A widely used alternative is the diabatic picture, brought about by a unitary transformation by which the non-adiabatic couplings vanish with a diabatic basis of states. The latter diabatic bases are not eigenstates of the electronic Hamiltonian, so that they are still coupled by electrostatic couplings, smaller and softer than non-adiabatic couplings. There is no single prescription to furnish a diabatic representation. Our approach to obtain diabatic states is described in more detail in the Appendix ~\ref{sec:adiabatic-diabatic}. The resulting diabatic electronic curves are shown in Fig.~\ref{fig:POC_shin-metiu}(b) (the avoided crossings become real crossings).

The electronic PECs and their ensuing vibrational states for the diabatic picture version of the Shin-Metiu system are illustrated in Fig.~\ref{fig:POC_shin-metiu}(b). After diabatization, the potential curves cross to each other and the vibrational eigenstates must be obtained from the nuclear equation that includes the new diabatic PECs. Then, above the energy of any energy crossing, the adiabatic and diabatic vibrational states differ. 
Consequently, in the diabatic picture the character exchange is now reduced to a minimum [see Fig.~\ref{fig:diabatic_states}(b)] but electronic diabatic states may differ from adiabatic ones
[compare Fig.~\ref{fig:diabatic_states}(a) with Fig.~\ref{fig:adiabatic_states}(a)].

\subsubsection{\bf{Electro-nuclear entanglement in the Shin-Metiu model}}

\paragraph{Adiabatic and diabatic pictures}

\begin{figure}
    \centering\includegraphics[scale=0.335]{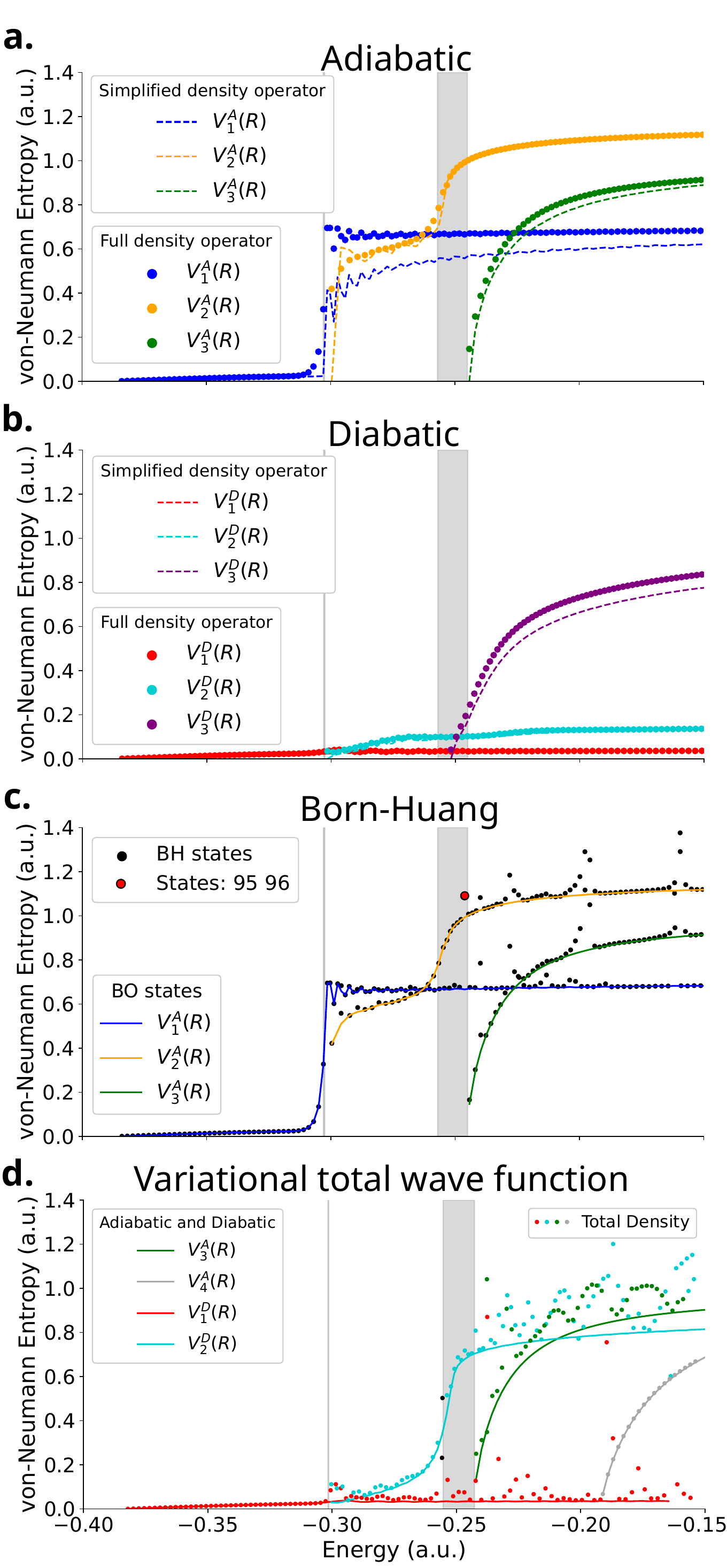}
    \caption{ Von Neumann entropy for the Shin-Metiu model using both adiabatic (a) and diabatic (b) pictures within the BO approximation, with a Born-Huang expansion (c) and the variational total wave function (d).   
    The entanglement content for each vibronic state $\phi_n (x;R) \chi_{n,m}(R)$ with $m=$0, max$_n$ for $n=1$ (blue), $n=2$ (yellow) and $n=3$ (green). The highest vibrational number max$_n$ corresponds to a vibronic energy close to $E=-0.15$ a.u. (a-b) Entanglement obtained from the full reduced density operator (dots) and from the simplified model (dashed lines) are plotted. The shaded gray regions indicate the zone of avoided crossings at $R_c^{(1,2)}$ (narrow) and $R_c^{(2,3)}$ (wide) (c)
    Entanglement of Born-Huang states (black dots) compared to that of BO adiabatic states. The big red dots highlight Born–Huang states 95 and 96 (see text), which the highest degree of mixing. (d) The entropy of vibronic states using a variational total eigenfunction of Eq.~\eqref{eq:hamiltoniano_shin_metiu} are plotted with dots. Solid lines represent the entropies of vibronic states that belong to the adiabatic or diabatic pictures chosen to reproduce the variational results at best (see text).
    }
\label{fig:entropy_shin-metiu}
\end{figure}

The von Neumann entropy for the lowest three potential energy curves of the Shin-Metiu model, in both adiabatic and diabatic pictures, is shown in Fig.~\ref{fig:entropy_shin-metiu}. 
These entanglement entropies are computed using the eigenvalues of both the full reduced density matrix and the reduced density obtained with the simplified model introduced in the previous section for H$^+_2$. 
The dimension of the simplified reduced density matrix is $m+1$, with $m$ being the label of the vibrational state or its number of nodes. 
\begin{figure*}[t]
\centering\includegraphics[scale=0.5]{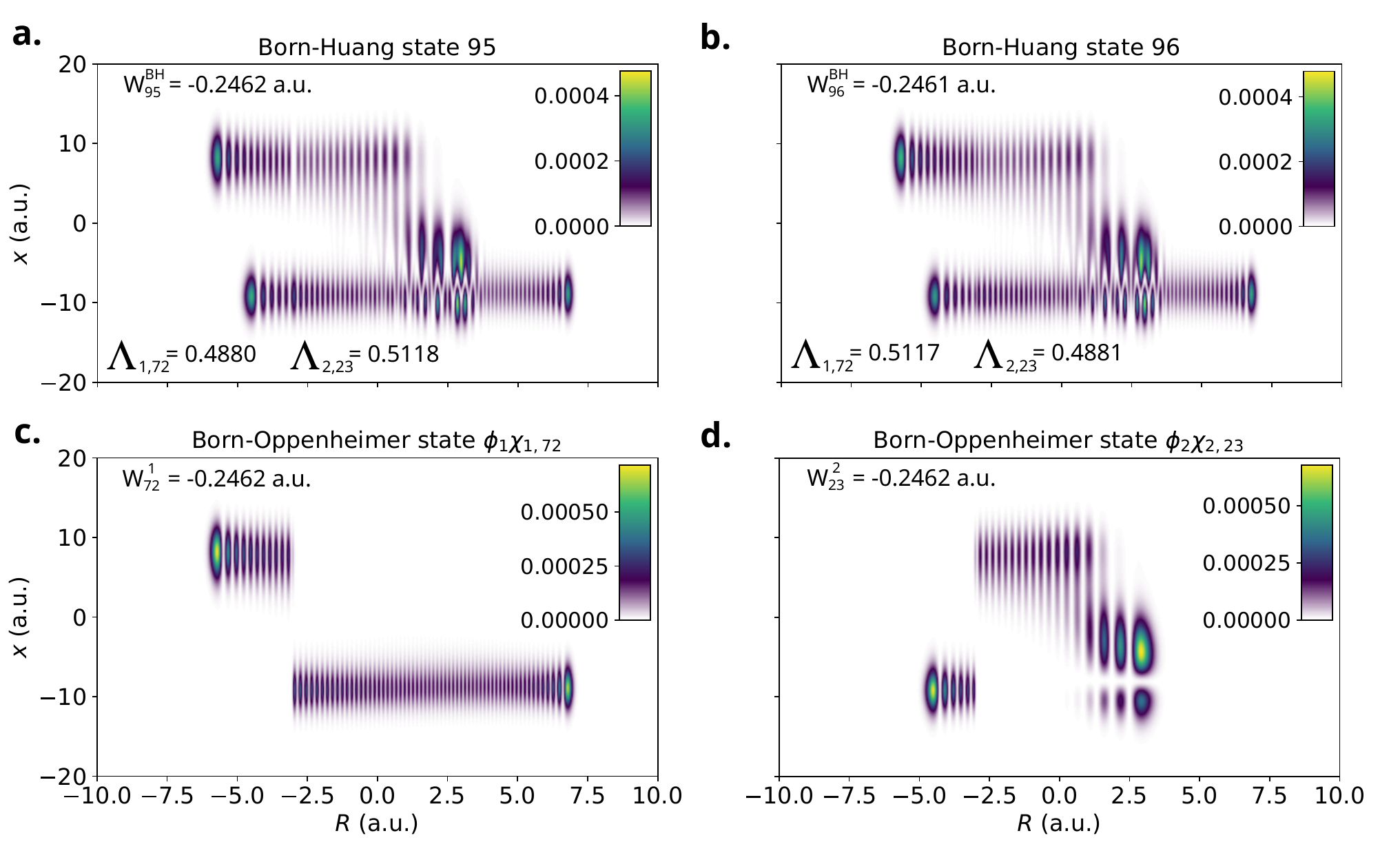}
    \caption{(a)-(b) Probability density of the total BH wave function for BH vibronic states 95 and 96. These two states are nearly degenerate, with energies $W_{95}^{\text{BH}} = -0.2462$ a.u. and $W_{96}^{\text{BH}} = -0.2461$ a.u., as indicated at the top of each panel. The leading coefficients in the BH expansion are also indicated as $\Lambda_{n,m}$.
    (c)-(d) Probability density of the total BO wave function of the adiabatic vibronic states $\phi_1 \chi_{1,72}$ and $\phi_2 \chi_{23}$ with a nearly degenerated vibronic energy $W^1_{72}=W^1_{23}=0.2462$. 
    }
\label{fig:schmidt_shin_metiu}
\end{figure*}

From Fig.~\ref{fig:entropy_shin-metiu}, we realize that the entanglement of the BO vibronic states in the adiabatic picture for PECs $V^A_1$ and $V^A_3$ follows the same pattern already observed in H$^+_2$ for the $1\sigma_g$ and $2\sigma_g$ states, respectively. 
The states in the PEC $V^A_1$ are not coupled below the lowest sharp avoided crossing at $E=-0.30$ a.u. and those in the PEC $V^A_3$ are barely coupled due to a wider avoided crossing with a weak coupling. The electronic WF of the electronic ground state $\phi_1(x;R)$ varies little from $R=-2.5$ to $R=6.0$ a.u., so that the two-point electronic overlap $S_1(R_k,R_\ell)$ remains close to unity for any pair. However, a sudden increase in the entanglement for $V^A_1$ occurs in the
energy region around the lowest avoided crossing at distance $R_c^{(1,2)}=-3.0$ a.u. and energy $E \sim -0.3$ a.u. In H$^+_2$ an asymptotic increase at $E=-0.5$ a.u. occurs at the dissociation threshold and the ensuing delocalization of the nuclear WF. 

However, in the Shin-Metiu model, the entanglement has a smooth behavior except for the regions of avoided crossings, where it shows the form of a step function. Note that in the avoided crossing zone there is a character exchange in the electronic WF; also the configuration space for the vibration changes. 
Both effects contribute to increasing the entanglement content by up to $0.6$ for BO vibronic states $\phi_1(x;R) \chi_{1,m}(R)$ in the PEC $V^A_1$ at the lowest avoided crossing, and from $0.6$ to $1.1$ for states $\phi_2(x;R) \chi_{2,m}(R)$ in the PEC $V^A_2$.
Above the energy $E=-0.30$ a.u. the entanglement for $V^A_1$ increases monotonically but very slowly. 
Taking into account the character exchange in the electronic WF, the overlap $S_1(R_k,R_\ell)$ now departs from unity for pairs of nuclear distances $\{R_k, R_\ell\}$ that meet $R_k < R_c^{(1,2)} < R_\ell$ or $R_\ell < R_c^{(1,2)} < R_k$. Apart from this exchange character, the electronic WFs change little when both distances $\{R_k, R_\ell\}$ lie above $R_c^{(1,2)}$ or below it. 
Then the entanglement above the avoided crossing remains similar in magnitude, as shown in Fig.~\ref{fig:entropy_shin-metiu}(a).
The BO states $\phi_2(x;R) \chi_{2,m}(R)$ in the PEC $V^A_2$ lying between the two avoided crossings [$E=-0.30$ and $E=-0.25$ a.u. approximately in Fig.~\ref{fig:POC_shin-metiu}(a)] increase rapidly their entanglement for the same reason. 
However, although the electronic WF $\phi_2(x;R)$ has a character exchange in $R_c^{(1,2)}$ the configurational space accessible to vibration that evaluates different overlaps $S_2(R_k,R_\ell)$ is much reduced. 

This case study indicates that the compactness in space (localization) of wave functions is not a sufficient condition to reduce the entanglement. Strong electron-nuclear correlations at a given geometry may produce a dramatic change in entanglement.  
In fact, even though BO PECs $V^A_1$ and $V^A_3$ are similar in shape, their lowest tens of vibronic states strongly differ in their entanglement, the former with a small electron-nuclear correlation that facilitates separability and the latter with an exchange of character across an avoided crossing.

Our simplified model for the reduced density matrix reproduces very well the entanglement behavior for the BO vibronic states $\phi_2(x;R) \chi_{2,m}(R)$ and $\phi_3(x;R)\chi_{3,m}(R)$, but for
$\phi_1(x;R) \chi_{1,m}(R)$ the simplification fails above the lowest avoided crossing at $E=-0.30$ a.u. We attribute this failure to insufficient sampling of nuclear points $\{R_k,R_\ell \} \in {\bf S}$, when the crossings are close to the turning points in the PECs. 
In this region nuclear WFs are characterized by the last wide oscillation with the highest amplitude. 
A single grid point located at the maximum cannot capture the abrupt changes in the electronic WF caused by a nearby avoided crossing.
However, the simplified model nicely follows the trend of the entanglement behavior calculated with the full grid.
  
In the diabatic picture [see Fig.~\ref{fig:entropy_shin-metiu}(b)], the magnitude of the von Neumman entanglement for the vibronic states in the PEC $V_1^D$ and $V_2^D$ is much reduced and with two small steps starting at the potential crossings. 
The diabatization procedure generates a smooth continuation of the electronic WFs $\phi_1(x, R)$ and $\phi_2(x, R)$ across the sharp avoided crossing $R_c^{(1,2)}$ and therefore the electronic overlaps $S_n(R_k,R_\ell)$ become closer to unity. In addition, WFs $\phi_2(x, R)$ and $\phi_3(x, R)$ are even smoother across the avoided crossing $R_c^{(2,3)}$. 
The entanglement of vibronic states in the diabatic PEC $V_3^D$ is not much affected and, indeed, only slightly reduced compared to the adiabatic picture. 
The only explanation comes from the fact that the diabatic electronic WF $\phi^D_3$ varies with nuclear geometry more than $\phi^D_1$ and $\phi^D_2$. 
This can be visualized in Fig.~\ref{fig:diabatic_states}(a)-(b); while the diabatic electronic WFs $\phi^D_1$ and $\phi^D_2$ remain localized and similar at two separate points $R_c^{1,2}=-3.0$ and  $R_c^{2,3}=1.2$, the WF $\phi^D_3$ changes noticeably. A similar pattern was found in H$^+_2$ for state $2\sigma_g$.

\paragraph{\label{sec:born-huang} Born-Huang expansion}
The Born–Huang expansion is believed to offer a systematic way to improve the total molecular state beyond the adiabatic BO approximation by proposing a total wave function as a superposition of electronic BO states in Eq.~\eqref{eq:Born-Huang_wave_function}. 
Non-adiabatic couplings between the different electronic states in the expansion are now explicitly included. The diagonalization of Eq.~\eqref{eq:BHnuclear}, in the FGH grid provides the total vibronic energies and states. 
Having access to Born-Huang solutions in a realistic molecular system may become prohibitive. 
This Shin-Metiu model allows for the accurate computation of all ingredients (electronic and vibrational wave functions and the whole set of non-adiabatic couplings). 

Although the expressions for the full density matrix in this basis [Eqs.~\eqref{eq:density_operator_R_BH} and \eqref{eq:density_operator_x_BH}] are more elaborate than those derived for pure BO states [Eqs.~\eqref{eq:density_operator_R} and \eqref{eq:density_operator_x}], its application is straightforward. Within the BH method the vibronic eigenstates cannot be attributed to a single potential curve (beyond BO there are no unambiguous PECs). Despite that, Fig.~\ref{fig:entropy_shin-metiu}(c) illustrates that the entanglement for most BH vibronic states aligns closely along one of the BO entropy values for $V^A_1$, $V^A_2$ and $V^A_3$. This happens when the mixing component of one BO wave function $\phi^A_n$ dominates. Nevertheless, there is always a BO superposition, so that the BH entanglement is always slightly above any BO counterpart. Incidentally, above $E=-0.30$ eV a couple of two nearly degenerated BO vibronic states from $V^A_1$ and $V^A_2$ with energies $E_{1,m_1} > E_{2,m_2}$,
interact with a (approximately) constant non-adiabatic coupling to produce a splitting in the energies and a new coupled BH vibronic eigenstate where the leading component is the $V^A_1$ component, then inherited by the entanglement content. The opposite also occurs for $E_{1,m_1} < E_{2,m_2}$ with a dominant contribution from $V^A_2$. 

However, Fig.~\ref{fig:entropy_shin-metiu}(c) also displays many other BH vibronic states whose entanglement deviates from the BO results and they correspond to states with significant mixing between multiple BO states, leading to an entropy value larger than that of the individual contributing BO states. These BH states typically appear in nearly degenerate doublets or triplets, a feature that reflects their origin in strongly coupled sets of BO vibronic states through the non-adiabatic couplings.
As an illustration, we highlight the case of BH vibronic states 95 and 96 (in the energy region of the second avoided crossing), which exhibit the highest mixture between two distinct BO vibronic states associated with different PECs $V^A_1$ and $V^A_2$. 
In Fig.~\ref{fig:schmidt_shin_metiu}(a)-(b), we plot the probability densities for the two BH vibronic states separated by a small energy splitting and with the superposition $0.4880 \phi_{1}\chi_{1,72}$ + $0.5118 \phi_{2}\chi_{2,23}$ for the lower BH vibronic state and $0.5117 \phi_{1}\chi_{1,72}$ + $0.4881 \phi_{2}\chi_{2,23}$, for the upper BH vibronic state.
Both states have almost the maximum  mixing 0.5/0.5 and their entropies are almost identical. 
These BH states are built from two energy-degenerated BO vibronic states, whose densities are plotted in Fig.~\ref{fig:schmidt_shin_metiu}(c)-(d), clearly inherited by the BH states. 
Above $E=-0.25$ a.u. there is a plethora of BH vibronic states with an entanglement that clearly departs from the BO discipline, and they correspond to other complex superpositions of three BO states, this time $\phi_2$ and $\phi_3$ having the leading role.   

\paragraph{Entanglement in the total molecular eigenfunction.} 
The total energies and WFs were also obtained by direct exact diagonalization of the Shin-Metiu Hamiltonian in Eq.~\eqref{eq:hamiltoniano_shin_metiu}. 
We use the same FGH grid basis mentioned above for the electronic and nuclear coordinates, in tensor product form.
The purpose of this calculation is to provide a reference to be compared to those previous solutions obtained within the adiabatic (both the BO and BH ansatzes) and diabatic frameworks.

%
\begin{figure}[t!]
    \centering\includegraphics[scale=0.43]{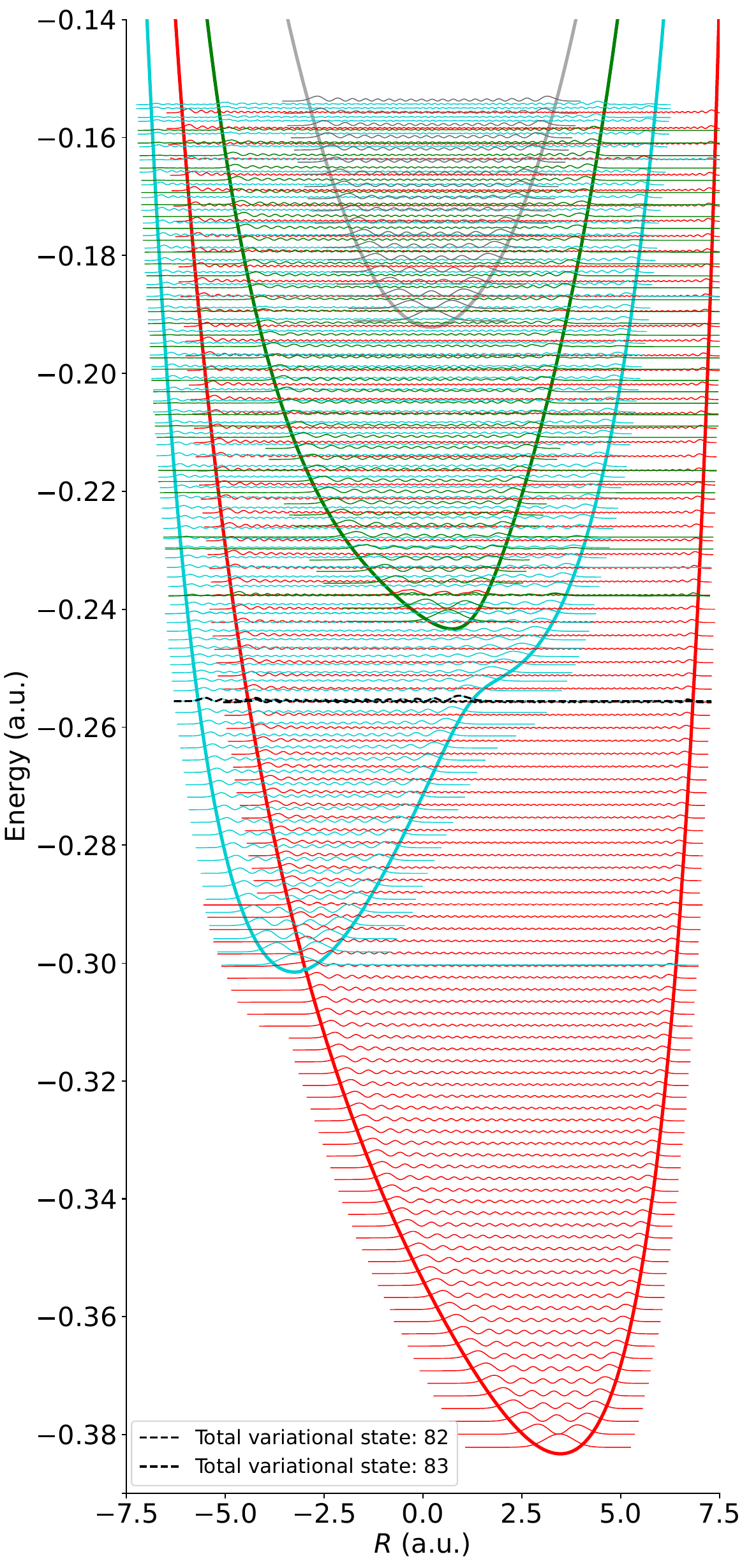}
    \caption{Nuclear probability densities for the vibronic states up to $E=-0.15$~a.u. coming from a variational total WF to solve the molecular Shin-Metiu model. The adiabatic and diabatic PECs up to the 4th state also are included as a reference to guide the eye, highlighting the classical turning points for each vibronic state. The nuclear densities of the states $82$ and $83$ with energy $E \sim 0.255$ are highlighted with black dashed lines (see also Appendix~\ref{Variational-solution}).
    }
    \label{fig:density_total}
\end{figure}

With the set of screening parameters used in this work, the Shin–Metiu model allows for the analysis of two distinct non-adiabatic regimes within the BO picture: a sharp avoided crossing (strong coupling) and a broad avoided crossing (weak coupling).
Fig.~\ref{fig:entropy_shin-metiu}(d) presents the von Neumann entropy of the variational vibronic states (with colored dots). The colored solid lines correspond to the entanglement computed for the adiabatic and diabatic states discussed previously [note that we have also included the entanglement corresponding to the adiabatic 4th state with PEC $V_4^{A}(R)$]. To clarify the observed trend, we performed the diabatization procedure only for the lowest sharp avoided crossing $V_1-V_2$ (entanglement of vibronic states in $V^D_1$ with red line and of vibronic states in PEC $V^D_2(r)$ with cyan line), but not for the broad avoided crossing $V_2-V_3$ (vibronic states in $V^D_2$ with blue line and vibronic states in PEC $V^D_3(R)$ with green line). 

To facilitate the analysis we plot the total nuclear propability densities for each total vibronic state $n$. They are obtained from the Schmidt decomposition \eqref{eq:SchmidtWF} of the total variational wave function \eqref{eq:totalvariationalWF}
\begin{equation}
    \rho_n^{nu}(R) = \int  |\Psi(x,R)|^2 dx =\sum_{i} \lambda_i^{n} |v_{i}^{n}(R)|^2.
\end{equation}
Fig.~\ref{fig:density_total} displays the nuclear probability densities of all non-BO variational vibronic states with energies below $E=-0.15$~a.u. (higher states are omitted for clarity). We realize that below the lowest avoided crossing at $E \sim -0.30$~a.u. the nuclear density of vibronic states follow the envelope of the adiabatic PEC $V^A_1(R)$, as expected, so do their related entropies of entanglement. Some near degenerate vibronic states around this crossing exhibit strong mixing.
Above this sharp avoided crossing, the nuclear densities mainly follow the diabatic picture with vibronic densities alternating between PECs $V^D_1(R)$ (red curve) and $V^D_2(R)$ (cyan curve).  
In contrast, the upper avoided crossing centered around $E~\sim -0.25$ a.u.largely preserves the adiabatic following of the nuclear densities within each respective PEC $V^A_2(R)$ and $V^A_3(R)$, although some degree of mixing always remains. The densities in solid gray lines in Fig.~\ref{fig:density_total} correspond to vibronic states that follow the adiabatic curve $V_4^{A}(R)$, which barely exhibit non-adiabatic mixing, 
so does the entanglement in Fig.~\ref{fig:entropy_shin-metiu}.\\

In Fig.~\ref{fig:entropy_shin-metiu}(d), the color of each point corresponds to the nuclear density displayed in Fig.~\ref{fig:density_total}.  
Although the total states are not purely adiabatic or diabatic, their classical turning points follow either adiabatic or diabatic PECs. The nuclear densities associated with red points mainly follow the diabatic curve $V_1^{D}(R)$, the cyan ones follow $V_2^{D}(R)$, and the green and gray points correspond to adiabatic curves $V_3^{A}(R)$ and $V_4^{A}(R)$, respectively. Appendix~\ref{Variational-solution} illustrates the comparison between two quasi-degenerated total variational eigenstates (labelled 82 and 83 in Fig.~\ref{fig:density_total}) with the corresponding (closer in energy) adiabatic and diabatic states.
The better comparison of the eigenstates with the diabatic states endorses our entanglement findings and supports the chemical intuition for a diabatization procedure when necessary.

Entanglement seems to be a very subtle property of multipartite WFs and it provides indeed a witness to their accuracy.
Our total variational eigenfunction (and its entanglement content) represents the molecular Shin-Metiu model at best, and provides a guidance on the quality of other approximated WFs. In terms of entanglement, the adiabatic BO approximation works fine for states far from avoided crossings (low mixing).
In this way entanglement content for vibronic states within the ground state below the crossing, and for all vibronic states within the 4th state is well represented by BO WFs. 
Instead, our entanglement results dictate that the vibronic states within the 1st and 2nd electronic states must be treated diabatically above the lowest avoided crossing, while in the region of the second avoided crossing and above, the vibronic states must be better treated adiabatically. 
It is worth noting that the BH expansion based on the BO states does not provide the correct diabatic trend when needed, thus indicating a slow convergence for the proper representation of the total WF.

\section{\label{sec:conclusions}Conclusions}

We have investigated the content of electro-nuclear entanglement in molecular wave functions. The quantum correlations between particles in a complex system like a molecule are already encoded within the Hamiltonian. A good representation of the correlations depends on the choice of a variational ansatz to solve the problem. Each variational molecular WF is then subjected to an alternative analysis in terms of quantum entanglement, which is not equivalent to correlation. 

Without loss of generality, we have analyzed the quantum entanglement between the electronic and nuclear motions present in variational WFs of molecular models in 1D. Although the BO approximation is sometimes invoked to claim that the electronic motion is separated from the nuclear one, the BO WF is indeed a kind of conditional product state $\Psi^{BO}({\bf r},{\bf R})$ = $\phi^{BO}({\bf r}; {\bf R}) \chi^{BO}( {\bf R})$ 
but not a fully separable product state $\phi({\bf r}) \chi({\bf R})$, so it is rigorously a quantum entangled state.  In this work, we have evaluated the electro-nuclear entanglement of BO WFs in H$^+_2$ for two lowest electronic states of symmetry $\sigma_g$ and their associated vibrational states. From this example, we learn that, contrary to the intuitive expectation that a larger configurational space (delocalization of the WF) leads to a larger entropy of entanglement, the variation of the electronic (nuclear) WF along the nuclear (electronic) distances plays a central role in the discrimination of molecular entanglement. 
Interestingly, our present results for the entanglement in 1D H$_2^+$ compare very well with those previously obtained with a non-BO ansatz to solve the 3D H$_2^+$ (states $1s\sigma_g$ and $3d\sigma_g$) \cite{Sanz2017}. Our results also show that the inclusion of molecular rotation does not change the entanglement content dramatically, indicating that the leading contribution comes in fact from the vibrational degrees of freedom.

Along this direction, we demonstrate that it is fair to consider the BO WF for the vibronic ground state to be almost separable and unentangled. However, the electro-nuclear entanglement increases with vibrational excitation. An extended excited-vibrational WF checks out more nuclear configurations in which differences in the electronic WF may appear. Since this proves to be a quantitative origin of entanglement, we have proposed a simplified model for the nuclear reduced density matrix
in which the information of the vibrational WF is encoded by retaining only the points with maximum amplitude, and we show that these nuclear positions are fairly enough to assess the variation of electronic WF. This simplification provides a computationally inexpensive method for estimating the entanglement as well as a simple perturbative approach for understanding the curvature behavior of entanglement at low vibrational excitations.
We also apply this framework to study abrupt changes in entanglement near avoided crossings in the PECs. The Shin-Metiu molecular model in 1D contains many of the expected ingredients in the photodynamics of molecules involving excited states: non-adiabatic transitions, adiabatic to diabatic pictures, etc. In the BO adiabatic picture, sharp avoided crossings produce rapid changes in entanglement as a result of abrupt modifications in the electronic WF (character exchange). In contrast, broad avoided crossings lead to smoother increases in entropy, as the electronic WF evolves smoothly across nuclear configurations. 

In the diabatic picture (where non-adiabatic couplings vanish) the change of entanglement across the real crossings is very much reduced. This result can be puzzling, but the entanglement of a bipartite system is indeed not unique; it depends on how the halfspaces are initially separated in the total WF. For example, in our previous work \cite{Sanz2017} we explored the entanglement of two coupled oscillators, a system that admits an exact solution in terms of normal modes. In the latter case, the total WF is a non-entangled direct product. In other local coordinates (which are related to the normal modes by a simple rotation), the WF becomes entangled. Similarly, the adiabatic and diabatic pictures are related by a basis transformation. Thus, the diabatic picture provides an ansatz for which electrons and nuclei display a much reduced entanglement, and we find that it eventually produces the correct representation for sharp avoided crossing (when compared to the total eigenfunction). 

If we move beyond the BO approximation, a proposal is to expand the total WF in terms of the BO adiabatic electronic states, and to include explicitly the non-adiabatic couplings (BH expansion). In general, the entropy of entanglement of a BH vibronic state closely follows that of the dominant contributing BO electronic state in the expansion. However, in certain cases,  significant mixing between BO vibronic states (localized on different PECs and quasi-degenerated in energy) leads to a noticeable increase in the entanglement of the BH vibronic state, beyond that of any BO contribution. In perspective, in the search for a maximum entanglement in vibronic states beyond the BO approximation, the entanglement content for each vibronic state might be analyzed individually. 

For a deep understanding of the results obtained within the BO approximation, we computed the entanglement with a variational eigenfunction of the total Hamiltonian. As mentioned above, from the entanglement results we conclude that the BO approximation and the BH expansion in terms of BO WFs do not adequately describe the correct WF in the presence of sharp avoided crossings, in favor of a diabatic representation. In contrast, broad avoided crossings with feeble couplings are reasonably described within the BO approximation and BH expansions. Our study here is performed on systems of reduced dimensionality, but we believe that our methods and conclusions remain valid for realistic molecules of higher dimensionality. \\

\begin{appendices}

\section{Adiabatic and diabatic states} \label{sec:adiabatic-diabatic}

There is no unique way to describe the diabatic representation of a coupled system. One approach involves the direct computation of the non-adiabatic coupling matrix elements (e.g., via the Hellmann–Feynman theorem), while another option is to perform successive unitary rotations in the vicinity of each avoided crossing \cite{Smith1969,Alden1982,Lecomte1979,Triana2022}. These two methods yield similar results and are theoretically equivalent \cite{Shu2022}. In the present work, we adopt the latter approach for convenience.

In the case of three adiabatic PECs, the corresponding matrix obtained by solving the Schr\"odinger equation [Eq.~\eqref{eq:molecular_hamiltonian}]. Within the BO approximation, it is block diagonal and it reads
\begin{equation}
   \mathbf{V}^{A}(R) =
        \begin{pmatrix}
            V_1^{A}(R) & 0 & 0 \\
            0 & V_2^{A}(R) & 0 \\
            0 & 0 & V_3^{A}(R)
        \end{pmatrix}.
\end{equation}
Since this matrix is already diagonal, there exists a unitary matrix $\mathbf{U}_T$ such that $\mathbf{V}^{D} = \mathbf{U}_T^\dagger \mathbf{V}^{A} \mathbf{U}_T$, where $\mathbf{V}^{D}$ contains the diabatic PECs. In the system studied here (with two avoided crossings located at $R_c^{(1,2)}$ and $R_c^{(2,3)}$), the diabatization procedure consists of a unitary transformation, in which a total rotation matrix is built from two consecutive rotations: $\mathbf{U}_T = \mathbf{U}_1 \mathbf{U}_2$. The first rotation, characterized by an angle $\theta(\mathbf{R})$, couples $V_1^{A}$ and $V_2^{A}$ near $R_c^{(1,2)}$, while the second rotation, characterized by an angle $\varphi(\mathbf{R})$, couples $V_2^{A}$ and $V_3^{A}$ near $R_c^{(2,3)}$.
Independent rotations are justified because of the spatial separation of the avoided crossings in our Shin-Metiu model.

\begin{figure}[t]
\centering\includegraphics[scale=0.4]{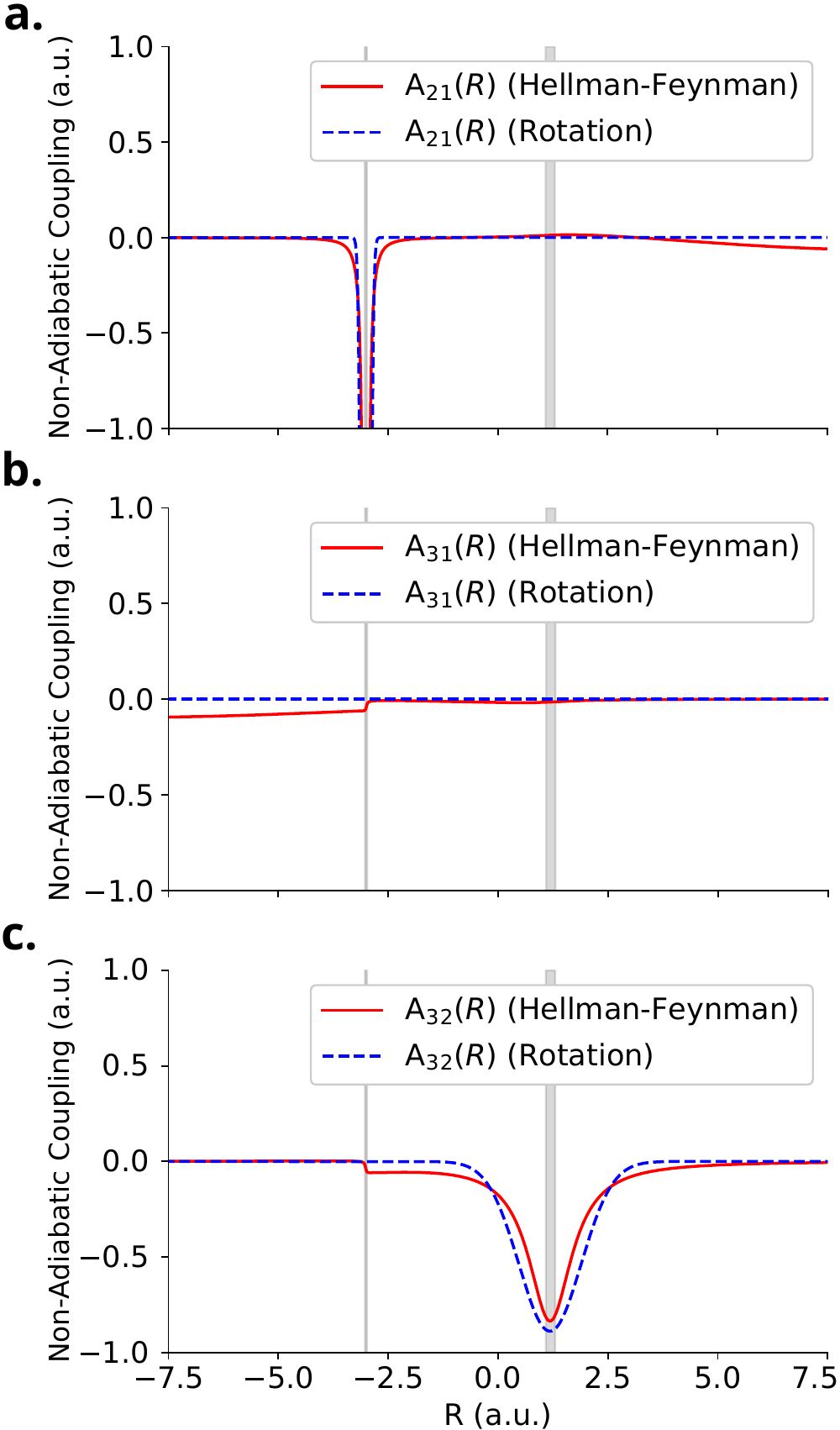}
    \caption{(a)-(c) Non-adiabatic coupling terms $A_{i,j}(R)$ computed using the Hellmann-Feynman theorem (red solid lines) and via rotation angles modeled with Gaussian distributions after Eq.~\eqref{eq:gaussian_distribution} (blue dashed lines). Vertical gray lines indicate the locations of the avoided crossings at $R_c^{(1,2)}$ and $R_c^{(2,3)}$, with the former being sharply localized and the latter in a wider region. 
    }
\label{fig:adiabatic-diabatic}
\end{figure}

The diagonal elements of the diabatic matrix are then given by
%
\begin{align}
    V_1^{D}(R) &= V_1^{A}(R) \cos^2\theta(R) + V_2^{A}(R) \sin^2\theta(R), \nonumber \\
    V_2^{D}(R) &= \left[ V_1^{A}(R) \sin^2\theta(R) + V_2^{A}(R) \cos^2\theta(R) \right] \cos^2\varphi(R) \nonumber \\
    & + V_3^{A}(R) \sin^2\varphi(R), \nonumber \\
    V_3^{D}(R) &= \left[ V_1^{A}(R) \sin^2\theta(R) + V_2^{A}(R) \cos^2\theta(R) \right] \sin^2\varphi(R) \nonumber \\ 
    & + V_3^{A}(R) \cos^2\varphi(R).
    \label{eq:dibatic_curves}
\end{align}
%
The off-diagonal elements (electrostatic couplings, since the diabatic states are not eigenstates of the electronic Hamiltonian) are
\begin{widetext}
\begin{align}
    V_{12}^{D}(R) &= V_{21}^{D}(R) = \frac{1}{2} \left( V_1^{A}(R) - V_2^{A}(R) \right) \sin 2\theta(R) \cos\varphi(R), \nonumber \\
    V_{13}^{D}(R) &= V_{31}^{D}(R) = \frac{1}{2} \left( V_1^{A}(R) - V_2^{A}(R) \right) \sin 2\theta(R) \sin\varphi(R), \nonumber \\
    V_{23}^{D}(R) &= V_{32}^{D}(R) = \frac{1}{2} \big[ V_1^{A}(R) \sin^2\theta(R) + V_2^{A}(R) \cos^2\theta(R) 
    - V_3^{A}(R) \big] \sin 2\varphi(R).
    \label{eq:electrostatic_couplings}
\end{align}
\end{widetext}

The angles $\theta(R)$ and $\varphi(R)$ can be modeled using error functions of the form
\begin{equation*}
    \theta(R),\ \varphi(R) \propto K\ \mathrm{erf}\left( \frac{R - R_{c}^{(i,j)}}{\Gamma} \right) + K_0,    
\end{equation*}
where $K$, $K_0$, and $\Gamma$ are parameters optimized for each angle to minimize the deviation from the conditions defined by Eq.~\eqref{eq:dibatic_curves}. The rotation angles must also satisfy the conditions below
\begin{align}
    \tan 2\theta(R)  &= \frac{V_{12}^{D}(R)}{V_1^{D}(R) - V_2^{D}(R)}, \nonumber\\
    \tan 2\varphi(R) &= \frac{V_{23}^{D}(R)}{V_2^{D}(R) - V_3^{D}(R)}.
    \label{eq:rotations_diabatic}
\end{align}

\section{\label{non-adiabatic_couplings} Non-adiabatic couplings}

\subsection{First order non-adiabatic couplings}

A direct method for computing the non-adiabatic couplings $\mathbf{A}_{n,n'}$, as defined in Eq.~\eqref{eq:non-adiabatic-couplings}, relies on the off-diagonal Hellmann-Feynman theorem~\cite{Balasubramanian1990}, that gives
\begin{equation}
    \mathbf{A}_{n,n'}(R) = \frac{i}{E_{n'}(R) - E_{n}(R)}
    \left< \phi_{n} \left| \partial_{\mathbf{R}} \hat{H}_e  \right| \phi_{n'}  \right>,
\end{equation}
for $n \neq n'$, and we take $\mathbf{A}_{n,n}(R) = 0$.  However, this method can become numerically unstable near degeneracies between two PECs. To overcome such issues, several alternative approaches have been proposed~\cite{Mandal2018}.

In our approach, we avoid direct computation of couplings by instead deriving an analytical expression based on the error function introduced above. This function governs the rotation angles $\theta(R)$ and $\varphi(R)$ between adiabatic and diabatic states as described in Appendix~\ref{sec:adiabatic-diabatic} and connects with the non-adiabatic couplings as
\begin{equation}
    \partial_R \theta(R) =  A_{1,2}(R), \ \ \partial_R \varphi(R) =  A_{2,3}(R). 
\end{equation}
Since the error function has an analytical derivative, the corresponding expressions are given by
\begin{equation}
     \partial_R\theta(R) \ \partial_R, \varphi(R) \propto  \frac{2 K}{\sqrt{\pi\Gamma^2}} \exp\left(- \frac{ \left(R - R_{c}^{(i,j)}\right)^2 }{\Gamma^2} \right).
     \label{eq:gaussian_distribution}
\end{equation}
Fig.~\ref{fig:adiabatic-diabatic} shows the agreement between the solution via Hellmann-Feynman and the couplings derived through angle rotations near the avoided crossings. Minor discrepancies are observed in $A_{1,2}(R)$ near the second avoided crossing at $R_c^{(2,3)}$, as seen in Fig.~\ref{fig:adiabatic-diabatic}(b). However, it becomes irrelevant since the vibrational wave functions $\chi_{2,m}(R)$ vanish for $R > 0$ (see Fig.~\ref{fig:POC_shin-metiu}).
Thus, this model can be effectively used for systems with well-separated avoided crossings, where the adiabatic to diabatic transformation can be represented by successive, independent rotations.

\subsection{Second order non-adiabatic couplings}

The second-order couplings $B_{n,n'}(R)$, involving second derivatives with respect to $R$, are typically more difficult to compute and often neglected. From the structure of Eq.~\eqref{eq:non-adiabatic-couplings}, the second-order term can be calculated using a completeness relation for the electronic states, with
\begin{equation}
    B_{n,n'} (R) = (\mathbf{A}\cdot \mathbf{A})_{n,n'}(R) = \sum_{l} A_{nl}(R) A_{ln'}(R), 
\end{equation}
Since we treat each avoided crossing independently, it suffices to construct two $2 \times 2$ matrices, to deal with the non-adiabatic coupling in the vicinity of $R_c^{(1,2)}$ and $R_c^{(2,3)}$, respectively.

For the sharp avoided crossing at $R_c^{(1,2)}$, the matrices take the form
\begin{equation}
    \mathbf{A}(R) = 
    \begin{pmatrix}
        0                   & \partial_R\theta(R) \\
        -\partial_R\theta(R)    & 0
    \end{pmatrix},
\end{equation}
and
\begin{equation}
    \mathbf{B}(R) = 
    \begin{pmatrix}
        -\left| \partial_R\theta(R) \right|^2    & \partial_R\theta(R) \\
        -\partial_R\theta(R)    & -\left| \partial_R \theta(R) \right|^2
    \end{pmatrix}.
\end{equation}
An equivalent representation holds for the second avoided crossing at $R_c^{(2,3)}$ using instead $\varphi(R)$.

\section{\label{Variational-solution} Total eigenfunctions vs. adiabatic and diabatic WFs in the Shin-Metiu model}

Here we include plots for the two quasi-degenerated variational states 
of the Shin-Metiu Hamiltonian (states 82 and 83 with energies $E=-0.2558$ and -0.2556 a.u.), with the corresponding (closest in energy) adiabatic states 
$\phi^A_1(x; R) \chi^A_{1,66}(R)$ ($E=-0.2561$ a.u.) and 
$\phi^A_2(x; R) \chi^A_{2,16}(R)$ ($E=-0.2553$ a.u.), respectively,  
and the closer diabatic states 
$\phi^D_1(x; R) \chi^D_{1,61}(R)$ ($E=-0.2558$ a.u.) and
$\phi^D_2(x; R) \chi^D_{2,21}(R)$ ($E=-0.2556$ a.u.), respectively.
In this energy region above the sharp avoided crossing, the WFs in the diabatic picture are those that better reproduce the total 
variational eigenfuntions. This WF similarity is then translated into their respective electron-nuclei entanglement.

\begin{figure*}[t]
\centering\includegraphics[scale=0.5]{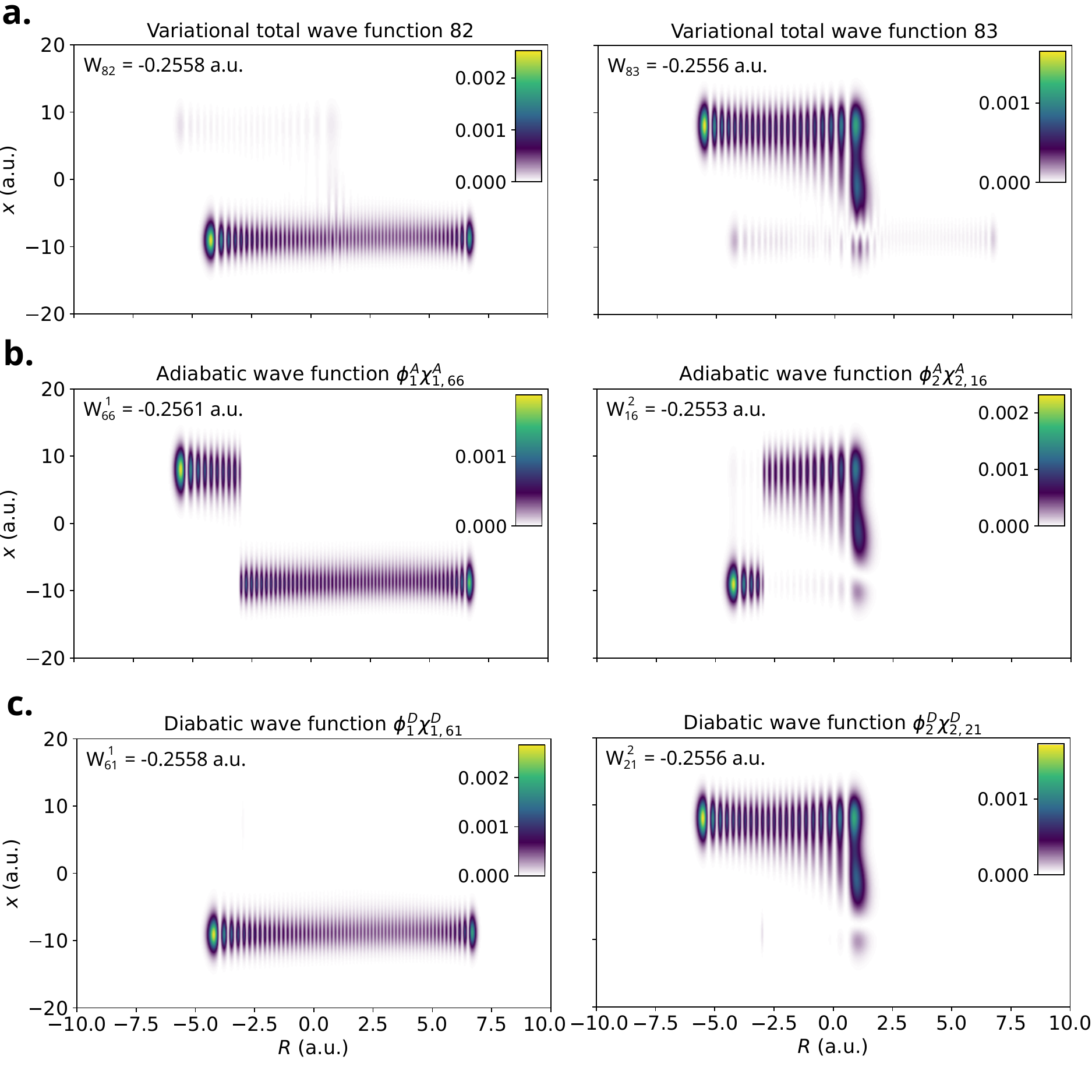}
    \caption{ (a) Variational total eigenfunctions of the Shin-Metiu Hamiltonian for two quasi-degenerated states ordered in energy with labels $82$ and $83$. (b)-(c) Adiabatic $\phi^A_n(x;R) \chi^A_{n,m} (R)$ ($m=66$ for $n=1$ and $m=16$ for $n=2$) and diabatic $\phi^D_n(x;R) \chi^D_{n,m} (R)$ ($m=61$ for $n=1$ and $m=21$ for $n=2$) WFs involving the lowest two PECs that are closer in energy to the total eigenfunctions shown in panel (a).
    }
\label{fig:total_wave_function}
\end{figure*}

\end{appendices}

\section{DATA AVAILABITY}
All the numerical results shown in this work can be  obtained straightforwardly from the formulas and parameters described in the text. However, the data are available from the authors upon reasonable request.

\begin{acknowledgments}

J.F.P.M and J.L.S-V acknowledge financial support from Vicerrectoría de Investigación at Universidad de Antioquia, CODI, Programática Project 2022-5357 and Estrategia de Sostenibilidad. 

\end{acknowledgments}

\bibliography{main}
\end{document}